\documentclass[aps,preprintnumbers,showpacs,twocolumn,superscriptaddress,floatfix,nofootinbib]{revtex4}
\usepackage{amsmath}
\usepackage{amssymb}
\usepackage{color}
\usepackage{dcolumn}
\usepackage{graphicx}
\usepackage{epstopdf}           
\usepackage{latexsym}
\usepackage{times}
\usepackage{ifpdf}

\input{epsf.sty}

\newcommand{\tg}{\tilde\gamma}
\newcommand{\tG}{\tilde\Gamma}
\newcommand{\tA}{\tilde A}

\newcommand{\dt}{(\partial_t - {\cal L}_\beta)\;}
\DeclareMathOperator{\tr}{\ensuremath{\mathrm{tr}}}

\newcommand{\bea}{\begin{eqnarray}}
\newcommand{\eea}{\end{eqnarray}}
\newcommand{\beq}{\begin{equation}}
\newcommand{\eeq}{\end{equation}}

\newcommand{\de}{\partial}
\newcommand{\dx}{\mathrm{d}}

\newcommand{\cf}{\textit{cf.}~}

\newcommand{\km}{{\rm km}}

\newcommand{\beqn}{\begin{eqnarray}}
\newcommand{\eeqn}{\end{eqnarray}}
\newcommand{\pa}{\partial}

\newcommand{\varep}{\varepsilon}

\begin{document}

\title{Binary neutron-star mergers with {\tt Whisky} and {\tt SACRA}: 
First quantitative comparison of results from independent
  general-relativistic hydrodynamics codes.}

\author{Luca Baiotti}
\affiliation{
Yukawa Institute for Theoretical Physics, Kyoto University, Kyoto,
606-8502, Japan
}
\affiliation{
Institute of Laser Engineering, Osaka University, Suita, 565-0871, Japan
}

\author{Masaru Shibata}
\affiliation{
Yukawa Institute for Theoretical Physics, Kyoto University, Kyoto, 606-8502, Japan
}

\author{Tetsuro Yamamoto}
\affiliation{
Yugen Club, Toyama, Shinjuku, Tokyo, 162-0052, Japan
}

\date{\today}

\begin{abstract}
  We present the first quantitative comparison of two independent
  general-relativistic hydrodynamics codes, the {\tt Whisky} code and
  the {\tt SACRA} code. We compare the output of simulations starting
  from the same initial data and carried out with the configuration
  (numerical methods, grid setup, resolution, gauges) which for each
  code has been found to give consistent and sufficiently accurate
  results, in particular in terms of cleanness of gravitational
  waveforms. We focus on the quantities that should be conserved during
  the evolution (rest mass, total mass energy, and total angular momentum)
  and on the gravitational-wave amplitude and frequency. We find that
  the results produced by the two codes agree at a reasonable level, with
  variations in the different quantities but always at better than
  about 10\%.
\end{abstract}

\maketitle

\section{Introduction}
\label{sec:Introduction}

Given the absence of astrophysically relevant exact solutions in
general relativity and the difficulty to compare results from
numerical-relativity codes with empirical observations (or
experiments), it is necessary to find alternative ways to assess the
capacity of existing codes to faithfully describe the physical
phenomena that they are supposed to simulate, and to check the
validity of their results. Among the strategies to achieve such a
reassuring confirmation, the most widely used are convergence tests
and checks of the violations of the physical constraints imposed by
the Einstein equations, in particular of the so-called Hamiltonian 
and momentum constraints, dictated by the choice of the 
Arnowitt-Deser-Misner (ADM) formalism as
a basis for numerical simulations (see
Eqs. (\ref{eq:einstein_ham_constraint}-\ref{eq:einstein_mom_constraints})
and, {\it e.g.}, Refs.~\cite{Baiotti:2009gk,Baiotti08,Yamamoto2008}). 
Another way to increase the
probability of having computer codes free from implementation-errors
and unaffected by possibly wrong and maybe hidden assumptions is the
comparison of the results of codes independently developed by separate
individuals or groups.

Since 2005, the year of the breakthrough in numerical
relativity~\cite{Pretorius:2005gq,Campanelli:2005dd,Baker:2005vv}
that made it possible to calculate the late inspiral, merger, and ringdown
of a black-hole binary system in full general relativity, and to
calculate the gravitational waves produced in the process, various
works~\cite{Hannam:2009hh,Baker-etal:2007a,Bruegmann:2006at} compared
the gravitational waveforms computed in vacuum simulations by several
codes. Their general conclusion is that the available codes give
consistent results (the difference among codes is smaller than the
estimated error within each code) and results that are good enough for
being of use in the quest for the detection of gravitational waves
through presently operating laser
interferometers~\cite{LIGO_web,VIRGO_web,GEO_web} or planned
detectors~\cite{Punturo:2010,ET_noise}.

In the present work, with in mind the goals delineated above, we perform and publish for the first time a
comparison between the results of two independent finite-difference codes solving the general-relativistic
hydrodynamics equations and the Einstein equations: the {\tt Whisky} 
code~\cite{Baiotti03a,Baiotti04,Giacomazzo:2007ti} and the {\tt SACRA}
code~\cite{Yamamoto2008}.
We include in the comparison also important quantities not directly related to the gravitational
waveforms, but, while giving a first glimpse of the comparison of the wave properties, we
postpone to a future article~\cite{Baiotti:2010b}, which may involve a larger number of codes, the detailed analysis of
the usefulness of the computed waveforms for current gravitational-wave detectors.

We also restrict our attention to the modeling of a single physical
system, the orbital inspiral of two neutron stars (NSs) in irrotational
configuration. This system is however one of the most promising
candidates for early detection of gravitational radiation and it is seen as the
most likely scenario leading to the formation of a black hole surrounded by a
massive torus with properties suitable for being the engine powering
short-hard gamma-ray bursts~\cite{Nakar:2007yr}.

We use a spacelike signature $(-,+,+,+)$ and a system of units in
which $c=G=M_\odot=1$ (unless explicitly shown otherwise for
convenience). Greek indices are taken to run from $0$ to $3$, Latin
indices from $1$ to $3$, and we adopt the standard convention for the
summation over repeated indices.

\section{Mathematical and Numerical Setup}
\label{sec:NumericalMethods}

All the details on the mathematical and numerical setup used by the
two codes have been discussed in depth in previous
works~\cite{Pollney:2007ss,Baiotti08,Yamamoto2008}. In what follows,
we limit ourselves to a brief overview, while spelling out the
differences between the two codes.

The differences in the implementation of the Einstein and
hydrodynamics equations between {\tt Whisky} and {\tt SACRA} are
summarized in Table~\ref{table:space-hydro-implem}.

\begin{table*}[t]
  \caption{Differences between {\tt Whisky} and {\tt SACRA} in the
    schemes for the evolution of the spacetime and of the
    hydrodynamics. See text for definitions and further explanations.}
\begin{ruledtabular}
\begin{tabular}{|lc|cc|c|}
 &  & {\tt Whisky} && {\tt SACRA} \\
\hline
  conformal factor $\phi$ & & evolve $\phi$ & & evolve $\chi\equiv e^{-2\phi}$ \\
\hline
  primitive matter variables && $\rho$, $v^i$, $\varep$ && $\rho$, $u_i$, $\varep$ \\
\hline
  evolved matter variables && $D$, $S_i$, $\tau$ && $D$, $S_i$, $E$ \\
\hline
  reconstructed matter variables & & primitive variables: $\rho$, $v^i$, $\varep$ 
  && $D$, $\tilde u_i\equiv S_i/D$, $\varep$  \\
\hline
  local Riemann solver & & Marquina flux formula & & central scheme
  (Kurganov and Tadmor) \\
\hline
  atmosphere treatment && constant rest-mass density &&
  exponentially decreasing rest-mass density \\
\end{tabular}
\end{ruledtabular}
\label{table:space-hydro-implem}
\end{table*}

\subsection{Evolution system for the fields}

We evolve the Einstein equations in the
Baumgarte-Shapiro-Shibata-Nakamura (BSSN) formalism~\cite{Nakamura87,
Shibata95, Baumgarte99, Alcubierre99d}. 

For the {\tt Whisky} simulations, all the equations discussed in this section and in the next
are solved using the \texttt{CCATIE} code, a three-dimensional finite-differencing code
based on the {\tt Cactus Computational Toolkit}~\cite{Goodale02a}. A
detailed presentation of the code and of its convergence properties
has been presented in~\cite{Pollney:2007ss}. For tests and details
on {\tt SACRA}, see instead~\cite{Yamamoto2008}.

In the BSSN formalism, the
spacetime is first decomposed into three-dimensional spacelike slices,
described by a metric $\gamma_{ij}$, an extrinsic curvature $K_{ij}$,
and the gauge functions $\alpha$ (lapse) and $\beta^i$ (shift) (see
Sec.~\ref{sec:Gauges} for details on how we treat gauges
and~\cite{York79} for a general description of the $3+1$ split). 
The standard 3+1 formulation is then modified by introducing different variables as
follows. The three-metric $\gamma_{ij}$ is conformally transformed via
\begin{equation}
  \label{eq:def_g}
  \phi = \frac{1}{12}\ln \det \gamma_{ij}, \qquad
  \tilde{\gamma}_{ij} = e^{-4\phi} \gamma_{ij},
\end{equation}
and the conformal factor $\phi$ (in {\tt Whisky}/{\tt CCATIE}) or a function of it
($\chi \equiv e^{-2\phi}$, in {\tt SACRA}) is evolved as an
independent variable, while $\tilde{\gamma}_{ij}$ is subject to the
constraint $\det \tilde{\gamma}_{ij} = 1$. The extrinsic curvature is
subjected to the same conformal transformation and its trace $K$ is 
evolved as an independent variable. That is, in place of
$K_{ij}$ we evolve:
\begin{equation}
  \label{eq:def_K}
  K \equiv \tr K_{ij} = \gamma^{ij} K_{ij}, \qquad
  \tilde{A}_{ij} = e^{-4\phi} (K_{ij} - \frac{1}{3}\gamma_{ij} K),
\end{equation}
with $\tr\tilde{A}_{ij}=0$. Finally, new evolution variables
\begin{equation}
  \label{eq:def_Gamma} \tilde{\Gamma}^i =
  \tilde{\gamma}^{jk}\tilde{\Gamma}^i_{jk} = -\tilde{\gamma}^{ij}_{~~,j}
\end{equation}
are introduced, defined in terms of the Christoffel symbols of
the conformal three-metric.

The Einstein equations specify a well-known set of evolution equations
for the listed variables. They are:
\begin{align}
  \label{eq:evolution}
  &\dt \tg_{ij} = -2 \alpha \tA_{ij}\,,  \\ \nonumber \\
  &\dt \phi = - \frac{1}{6} \alpha K\,,~
{\rm or}~~~
\dt \chi =  \frac{1}{3} \alpha \chi K\,,
 \\ \nonumber \\ 
  &\dt \tA_{ij} = e^{-4\phi} [ - \mathcal{D}_i \mathcal{D}_j \alpha 
   + \alpha (R_{ij} - 8 \pi \mathcal{S}_{ij}) ]^{TF} \nonumber\\
   & \hskip 2.0cm + \alpha (K \tA_{ij} - 2 \tA_{ik} \tA^k{}_j), \\
  &\dt K  = - \mathcal{D}^i \mathcal{D}_i \alpha \nonumber \\
   & \hskip 2.0cm + \alpha \Big [\tA_{ij} \tA^{ij} + \frac{1}{3} K^2 + 
  4\pi (\rho_{_{\rm  ADM}}+\mathcal{S})\Big ], \\ \nonumber \\
& 
  \partial_t \tG^i  = \tilde\gamma^{jk} \partial_j\partial_k \beta^i
    + \frac{1}{3} \tilde\gamma^{ij}  \partial_j\partial_k\beta^k
    + \beta^j\partial_j \tilde\Gamma^i
   - \Tilde\Gamma^j \partial_j \beta^i \nonumber \\
   & \hskip 1.0cm
   + \frac{2}{3} \tilde\Gamma^i \partial_j\beta^j 
   - 2 \tilde{A}^{ij} \partial_j\alpha
   + 2 \alpha ( 
   \tilde{\Gamma}^i{}_{jk} \tilde{A}^{jk} + 6 \tilde{A}^{ij}
   \partial_j \phi \nonumber\\
   & \hskip 1.0cm - \frac{2}{3} \tg^{ij} \partial_j K - 8 \pi \tg^{ij} \mathcal{S}_j),
\end{align}
where $R_{ij}$ is the three-dimensional Ricci tensor, $\mathcal{D}_i$ the
covariant derivative associated with the three metric $\gamma_{ij}$,
``TF'' indicates the trace-free part of tensor objects, $\mathcal{S} \equiv \gamma^{ij}\mathcal{S}_{ij}$, and $
{\rho}_{_{\rm ADM}}$, $\mathcal{S}_j$, and $\mathcal{S}_{ij}$ are the matter source terms
defined as
\begin{align}
\rho_{_{\rm ADM}}&\equiv n_\alpha n_\beta T^{\alpha\beta}, \nonumber \\ 
\mathcal{S}_i&\equiv -\gamma_{i\alpha}n_{\beta}T^{\alpha\beta}, \\
\mathcal{S}_{ij}&\equiv \gamma_{i\alpha}\gamma_{j\beta}T^{\alpha\beta}, \nonumber
\end{align}
where $n_\alpha\equiv (-\alpha,0,0,0)$ is the future-pointing
four-vector orthonormal to the spacelike hypersurface and
$T^{\alpha\beta}$ is the stress-energy tensor for a perfect fluid
({\it cf.} Eqs.~[\ref{hydro eqs})]. The Einstein equations also lead to
a set of physical constraint equations that are satisfied within each
spacelike slice,
\begin{align}
  \label{eq:einstein_ham_constraint}
  \mathcal{H} &\equiv R^{(3)} + K^2 - K_{ij} K^{ij} - 16\pi\rho_{_{\rm ADM}} = 0, \\
  \label{eq:einstein_mom_constraints}
  \mathcal{M}^i &\equiv \mathcal{D}_j(K^{ij} - \gamma^{ij}K) - 8\pi \mathcal{S}^i = 0,
\end{align}
which are usually referred to as Hamiltonian and momentum constraints, respectively.
Here $R^{(3)}=R_{ij} \gamma^{ij}$ is the Ricci scalar on a
three-dimensional timeslice. Our specific choice of evolution
variables introduces five additional constraints,
\begin{align}
  \det \tilde{\gamma}_{ij} & = 1, 
    \label{eq:gamma_one}\\
  \tr \tilde{A}_{ij} & = 0,
    \label{eq:trace_free_A}\\
  \tilde{\Gamma}^i & = \tilde{\gamma}^{jk}\tilde{\Gamma}^i_{jk}.
   \label{eq:Gamma_def}
\end{align}
Our codes actively enforce the algebraic
constraints~(\ref{eq:gamma_one}) and~(\ref{eq:trace_free_A}).  
Specifically, after every time evolution, we perform a reset as follows:
\begin{align}
 \tilde \gamma_{ij}
&\rightarrow [{\rm det}(\tilde \gamma_{ij})]^{-1/3}
\tilde \gamma_{ij},\\
 \tilde A_{ij} &\rightarrow
[{\rm det}(\tilde \gamma_{ij})]^{-1/3} \tilde A_{ij}
-{1 \over 3}\tilde \gamma_{ij}{\rm Tr}(\tilde A_{ij}), \\
K &\rightarrow K + {\rm Tr}(\tilde A_{ij}). 
\end{align}
In {\tt SACRA}, the additional resetting
\begin{align}
e^{-2\phi} &\rightarrow  [{\rm det}(\tilde \gamma_{ij})]^{-1/6} e^{-2\phi}
\end{align}
is performed. We note that in these adjustments $\gamma_{ij}$ and
$K_{ij}$ are unchanged.

The remaining constraints, $\mathcal{H}$, $\mathcal{M}^i$,
and~(\ref{eq:Gamma_def}), are not actively enforced and can be used
as monitors of the accuracy of our numerical solution.
See~\cite{Alcubierre02a} for a more comprehensive discussion of
these points.

\subsection{Gauges}
\label{sec:Gauges}

We specify the gauge in terms of the standard ADM lapse
function, $\alpha$, and shift vector, $\beta^i$~\cite{misner73}.
We evolve the lapse according to the ``$1+\log$'' slicing
condition~\cite{Bona94b}:
\begin{equation}
  \partial_t \alpha - \beta^i\partial_i\alpha 
    = -2 \alpha K.
  \label{eq:one_plus_log}
\end{equation}
The shift is evolved using the hyperbolic $\tilde{\Gamma}$-driver
condition~\cite{Alcubierre02a}
\begin{eqnarray}
\label{shift_evol}
  \partial_t \beta^i - \beta^j \partial_j  \beta^i & = & \frac{3}{4} B^i\,,
  \\
  \partial_t B^i - \beta^j \partial_j B^i & = & \partial_t \tilde\Gamma^i 
    - \beta^j \partial_j \tilde\Gamma^i - \eta B^i\,,
\end{eqnarray}
where $\eta$ is a parameter which acts as a damping coefficient. 
We set it to be constant and $\approx 3/M_{\rm b}$, where $M_{\rm b}$ is
the baryon mass of one of the stars (for the simulations made with
{\tt Whisky} in the present work, the results do not change
appreciably if $\eta$ is changed at least within a factor $2$ of the
above value). The advection terms on the right-hand sides of these
equations have been suggested in~\cite{Baker05a, Baker:2006mp,
Koppitz-etal-2007aa}.

\subsection{Apparent horizons and gravitational waves}

After the merger, the apparent horizon (AH) formed during the
simulation is located every few timesteps during the evolution. In
{\tt Whisky} this computation is performed both with the {\tt
AHFinderDirect} code of \cite{Thornburg2003:AH-finding_nourl,Thornburg95} and in the
isolated and dynamical-horizon
frameworks~\cite{Ashtekar99a,Ashtekar00a,Ashtekar01a,Ashtekar-etal-2002-dynamical-horizons,Dreyer-etal-2002-isolated-horizons}. In
{\tt SACRA} the AH is located as reported in~\cite{Yamamoto2008}.\\

For the results reported in the present work, both codes extract the gravitational 
waves using the Newman-Penrose formalism, which provides a
convenient representation for a number of radiation-related quantities
as spin-weighted scalars. In particular, the curvature scalar
\begin{equation}
  \Psi_4 \equiv -C_{\alpha\beta\gamma\delta}
    n^\alpha \bar{m}^\beta n^\gamma \bar{m}^\delta
  \label{eq:psi4def}
\end{equation}
is defined as a particular component of the Weyl curvature tensor
$C_{\alpha\beta\gamma\delta}$ projected onto a given null frame
$\{\boldsymbol{l}, \boldsymbol{n}, \boldsymbol{m},
\bar{\boldsymbol{m}}\}$ and can be identified with the gravitational
radiation if a suitable frame is chosen at the extraction radius. In
practice, we define an orthonormal basis in the three-space
$(\hat{\boldsymbol{r}}, \hat{\boldsymbol{\theta}},
\hat{\boldsymbol{\phi}})$, centered on the Cartesian origin and
oriented with poles along $\hat{\boldsymbol{z}}$. The normal to the
slice defines a timelike vector $\hat{\boldsymbol{t}}$, from which we
construct the null frame
\begin{equation}
   \boldsymbol{l} = \frac{1}{\sqrt{2}}(\hat{\boldsymbol{t}} - \hat{\boldsymbol{r}}),\quad
   \boldsymbol{n} = \frac{1}{\sqrt{2}}(\hat{\boldsymbol{t}} + \hat{\boldsymbol{r}}),\quad
   \boldsymbol{m} = \frac{1}{\sqrt{2}}(\hat{\boldsymbol{\theta}} - 
     {\mathrm i}\hat{\boldsymbol{\phi}}) \ .
\end{equation}
We then calculate $\Psi_4$ via a reformulation of (\ref{eq:psi4def}) 
in terms of ADM variables on the slice~\cite{Shinkai94}:
\begin{equation}
  \Psi_4 = C_{ij} \bar{m}^i \bar{m}^j,  \label{eq:psi4_adm}
\end{equation}
where
\begin{equation}
  C_{ij} \equiv R_{ij} - K K_{ij} + K_i{}^k K_{kj} 
    - {\rm i}\epsilon_i{}^{kl} \nabla_l K_{jk}
\end{equation}
and $\epsilon_{ijk}$ is the Levi-Civita symbol.
The gravitational-wave polarization amplitudes $h_+$ and $h_\times$
are then related to $\Psi_4$ by time
integrals~\cite{Teukolsky73}:
\begin{equation}
\ddot{h}_+ - {\rm i}\ddot{h}_{\times}=\Psi_4 \ ,
\label{eq:psi4_h}
\end{equation}
where the double overdot stands for the second-order time derivative.
Caution should be taken when performing such integrals~\cite{Baiotti:2008nf}.

For the extraction of the gravitational-wave signal, both codes also implement 
an independent method, which is based on the measurements
of the nonspherical gauge-invariant metric perturbations of a background 
spacetime~\cite{Moncrief74}. The wave data obtained in this way give results compatible with
the ones obtained with the Newman-Penrose formalism and are not reported here.

\subsection{Evolution system for the matter}
\label{hd_eqs}

Both codes adopt a \textit{flux-conservative} formulation of the hydrodynamics
equations~\cite{Marti91,Banyuls97,Ibanez01}, in which the set of
conservation equations for the stress-energy tensor $T^{\mu\nu}=\rho h u^\mu u^\nu+p g^{\mu\nu}$ and
for the matter current density $J^\mu=\rho u^{\mu}$ (see below for definitions), namely
\begin{equation}
\label{hydro eqs}
\nabla_\mu T^{\mu\nu} = 0\;,\;\;\;\;\;\;
\nabla_\mu J^\mu = 0\, ,
\end{equation}
is written in a hyperbolic, first-order, flux-conservative form of
the type
\begin{equation}
\label{eq:consform1}
\partial_t {\mathbf q} + 
        \partial_i {\mathbf f}^{(i)} ({\mathbf q}) = 
        {\mathbf s} ({\mathbf q})\ ,
\end{equation}
where ${\mathbf f}^{(i)} ({\mathbf q})$ and ${\mathbf s}({\mathbf q})$
are the flux vectors and source terms, respectively~\cite{Font03}. Note
that the right-hand side (the source terms) does not depend
on derivatives of the stress-energy tensor. Furthermore,
while the system (\ref{eq:consform1}) is not strictly hyperbolic,
strong hyperbolicity is recovered in a flat spacetime, where ${\mathbf s}
({\mathbf q})=0$.

The \textit{primitive} hydrodynamical variables are the rest-mass
density $\rho$, the specific internal energy $\varepsilon$ measured in
the rest-frame of the fluid, and the fluid three-velocity (defined as
$v^i=u^i/W+\beta^i/\alpha$ (contravariant components) in {\tt Whisky} and as $u_i$ (covariant components) in 
{\tt SACRA}, where $u^{\mu}$ is the four-velocity measured by a local
zero--angular-momentum observer; {\tt SACRA} defines contravariant components of the three-velocity
as $V^i=u^i/u^0$). The Lorentz factor is defined as 
\begin{eqnarray}
W &\equiv & \alpha u^0=(1+\gamma^{ij}u_i u_j)^{1/2} \nonumber \\
&=& (1-\gamma_{ij}v^i v^j)^{-1/2}. 
\end{eqnarray}
There is then an equation of state (EoS) relating
pressure, rest-mass density and internal-energy density.

Following~\cite{Banyuls97}, in order to write system (\ref{hydro eqs})
in the form of system (\ref{eq:consform1}), the primitive variables
are mapped to a set of \textit{conserved} variables
\mbox{${\mathbf q} \equiv (D, S_i, E)$} via the relations
\vspace{-0.2 cm}
\begin{eqnarray}
  \label{eq:prim2con}
   D &\equiv& \sqrt{\gamma}W\rho = e^{6\phi}W\rho\ , \nonumber\\
   S_i &\equiv& D \tilde u_i = \sqrt{\gamma} \rho h W^2 v_i\\ 
   E &\equiv& \sqrt{\gamma}\left( \rho h W^2 - p\right) \equiv \tau+D \equiv D\tilde e, \nonumber
\end{eqnarray}
where $h \equiv 1 + \varep + p/\rho$ is the specific enthalpy, 
$\tilde u_i\equiv h u_i$ is the specific momentum, and 
$\tilde e\equiv h W - p/(\rho W)$ is the specific energy. 

In this approach, all variables ${\bf q}$ are represented on the
numerical grid by cell-integral averages. The functions that the
variables ${\bf q}$ represent are then {\it reconstructed} within each
cell, usually by piecewise polynomials, in a way that preserves
conservation of the variables ${\bf q}$~\cite{Toro99}. This operation
produces two values at each cell boundary, which are then used as
initial data for the local Riemann problems, whose (approximate)
solution gives the fluxes through the cell boundaries. A
method-of-lines approach~\cite{Toro99}, which reduces the partial
differential equations~\eqref{eq:consform1} to a set of ordinary
differential equations that can be evolved using standard numerical
methods, such as Runge-Kutta or the iterative Cranck-Nicholson
schemes~\cite{Teukolsky00,Leiler_Rezzolla06}, is used to update the
equations in time (see~\cite{Baiotti03a} for further details).
Here, we employ the $4^{\rm th}$-order Runge-Kutta method (see below). 

Various reconstruction methods are implemented in {\tt Whisky} and
{\tt SACRA}, but here we always use the piecewise parabolic method
(PPM)~\cite{Colella84}.
Both codes implement the scheme of Kurganov-Tadmor \cite{Tadmor00}
(which is a variation of the HLLE approximate Riemann solver~\cite{Harten83}), but
{\tt Whisky} gets better results employing the Marquina flux
formula~\cite{Aloy99b} (see~\cite{Baiotti03a,Baiotti04} for a more
detailed discussion). A comparison among different numerical methods
in binary-evolution simulations was reported
in~\cite{Baiotti08,Giacomazzo:2009mp}.\\

There are differences between {\tt Whisky} and {\tt SACRA} in several
implementation choices. In {\tt Whisky}, the variables whose evolution
is computed are $D$, $S_i$, and $\tau \equiv E-D$. {\tt SACRA} adopts
as evolution variables $D$, $S_i$, and $E$. Furthermore, the PPM
reconstruction is performed by {\tt SACRA} on the variables $\rho$,
$\tilde u_i=S_i/D$, and $\varepsilon$, while {\tt Whisky} reconstructs
the primitive variables $\rho$, $v^i$, and $\varep$.

Other differences are present in the conversion from the evolved
conservative variables back to the primitive variables, which are used
to calculate the fluxes and the source terms of the equations.  Such a
conversion cannot be given in an analytical closed form (except in
certain special circumstances).

{\tt Whisky} implements the following procedure to do the
conversion. One writes an equation for the pressure
\begin{equation}
  \label{eq:pressure1}
  p - \bar{p}\big[\rho({\bf q},p),\varep({\bf q},p)\big] = 0\ ,
\end{equation}
where $p$ is the value of the pressure to be found and
$\bar{p}[\rho({\bf q},p),\varepsilon({\bf q},p)]$ is the pressure as
obtained through the EoS in terms of the updated conserved variables
${\bf q}$ and of $p$ itself. This is done by inverting
\eqref{eq:prim2con} to express $\rho$ and $\varep$ in terms of the
conserved variables and of the pressure only:
\begin{eqnarray}
  \label{eq:press1gives}
  \rho & = & \frac{D}{\tau + p + D}
  \sqrt{ (\tau + p + D)^2 - S^2 }\ , \\
\nonumber\\
  \varepsilon 
& = & D^{-1} \left[ \sqrt{ (\tau +
      p + D)^2 - S^2 } - p \overline{W} - D
  \right]\ ,
\end{eqnarray}
where
\begin{equation}
  \overline{W} = \frac{\tau + p + D}{\sqrt{
      (\tau + p + D)^2 - S^2 }}
\end{equation}
is the Lorentz factor, expressed in terms of the conserved variables, and 
\begin{equation}
  \label{eq:s2}
  S^2 \equiv \gamma^{ij}S_iS_j\ .
\end{equation}
Then \eqref{eq:pressure1} is solved numerically. In {\tt Whisky} we
use a Newton-Raphson root finder, for which we
need the derivative of the function with respect to the dependent
variable, {\it i.e.} the pressure. This is given by
\begin{eqnarray}
  \label{eq:df}
\frac{\dx}{\dx p} \Big\{  p &-& \bar{p}\big[\rho({\bf q},p),\varepsilon({\bf q},p)\big]\Big\}
\nonumber \\
 &=& 1 - \frac{\partial \bar{p}(\rho, \varepsilon)}{\partial \rho}\frac{\partial
  \rho}{\partial p} - \frac{\partial \bar{p}(\rho,\varepsilon)}{\partial
  \varepsilon}\frac{\partial \varepsilon}{\partial p}\ , 
\end{eqnarray}
where 
\begin{eqnarray}
  \label{eq:df2}
  \frac{\partial \rho}{\partial p} & = & \frac{D
      S^2}{\sqrt{(\tau + p + D)^2 -
      S^2}(\tau + p + D)^2}, \\
  \frac{\partial \varepsilon}{\partial p} & = & \frac{p
      S^2}{\rho\left[(\tau + p + D)^2 -
      S^2\right](\tau + p + D)},
\end{eqnarray}
and where $\partial \bar{p}/\partial \rho$ and $\partial
\bar{p}/\partial \varepsilon$ are given by the EoS. Once the pressure is
found, the other variables follow simply.

In {\tt SACRA}, the conversion is performed in the following way. 
From the normalization relation $u^{\mu}u_{\mu}=-1$, $W$ is expressed 
in terms of $h$ and of the evolved values of $\tilde u_i(=S_i/D)$ 
and $\gamma^{ij}$:
\beqn
W^2=1 + {\gamma^{ij} \tilde u_i \tilde u_j \over h^2}. \label{eq-Wh}
\eeqn
For the EoSs chosen in the present work, $h$ is regarded as a function
of $W$ for the evolved values of $D$, $\tilde e$, and $\phi$ because
of the relation $\tilde e = h W - p e^{6\phi}/D$ and of the fact that
$p$ is written as a function of $h$, $\rho(=De^{-6\phi}/W)$, and $W$
as $p=p(h, \rho)=p(h, W)$. Substituting the resulting relation for
$h=h(W)$ into Eq.~(\ref{eq-Wh}), we obtain a one-dimensional algebraic
equation for $W$ both for the $\Gamma$-law EoS and the piecewise-polytropic 
EoS (see Sec.~\ref{sec:EoS}). We solve this derived equation using the
Newton-Raphson method, for which we need to take a derivative of the
equation $F(W)=0$ with respect to $W$. This is rather straightforward, 
and straightforward is also the determination of
the variables $h$, $\varepsilon$, and $P$, once the
equation for $W$ is solved.

\subsubsection{Treatment of the atmosphere}
\label{sec:atm-treatment}

At least mathematically, the region outside our initial stellar models
is assumed to be perfect vacuum. Independently of whether this
represents a physically realistic description of a compact star, the
vacuum represents a singular limit of any conservative scheme for
hydrodynamical evolution and must be treated artificially. Both codes
follow a standard approach in computational fluid-dynamics, that is
the addition of a tenuous ``atmosphere'' filling the computational
domain outside the star.

Of course, the density of the atmosphere should be as small as
possible, in order to avoid spurious effects. The evolution of the
hydrodynamical equations in grid zones where the atmosphere is present
is the same as the one used in the bulk of the flow. When the rest-mass 
density in a grid zone falls below the threshold set for the atmosphere,
that grid zone is not updated in time and the values of its rest-mass
density, internal-energy density, and velocity are set to those of the atmosphere.

Both codes treat the atmosphere as a zero--coordinate-velocity perfect fluid governed by a
polytropic EoS with the same adiabatic index used for the bulk
matter, or, in case of the piecewise-polytropic EoS (see Sec.~\ref{sec:EoS}), the
same adiabatic index as the one used in the outer parts of the
star\footnote{In this case, also the polytropic constant $K$ for the 
atmosphere is chosen to be the same as the one in the
outer parts of the star.}. However, the values of the rest-mass density
assigned to the atmosphere are different.  In {\tt Whisky},
the rest-mass density is set to be constant and several ($10$ in the
present simulations) orders of magnitude smaller than the initial
maximum rest-mass density $\rho_{\rm max}$~\cite{Baiotti03a,Baiotti04,Baiotti08}.

In {\tt SACRA} the rest-mass density is assigned as
\beqn
\rho=
\left\{
\begin{array}{ll}
\rho_{_{\rm atmo}} & r \leq r_{_0}, \\
\rho_{_{\rm atmo}}e^{1-r/r_0} & r > r_{_0}, 
\end{array}
\right.
\eeqn
where $\rho_{_{\rm atmo}}=\rho_{_{\rm max}} \times 10^{-9}$ is
chosen. $r_{_0}$ is a coordinate radius of about $10$--$20M_{_{\rm ADM}}$, where $M_{_{\rm ADM}}$
is the ADM mass of the system. In both codes, also the internal-energy density $\varepsilon$ is then 
recomputed from $\rho$ according to the polytropic EoS.

For both codes, with such a choice of parameters, the rest mass of the atmosphere is at least a
factor $10^{-5}$ smaller than the rest mass of the NSs. Thus, spurious effects due to the
presence of the atmosphere, such as accretion of the atmosphere onto
the NSs and the black hole, the resulting dragging effect against orbital
motion, gravitational effects, and effects on the formation and
dynamics of the disk around the merged object play a negligible role
in the present context.

\begin{table*}[ht]
  \caption{Differences in the implementation of the AMR of {\tt SACRA}
  and {\tt Whisky}. See text for definitions and further explanations.}
\begin{ruledtabular}
\begin{tabular}{|l|c|c|}
  & {\tt Whisky} & {\tt SACRA} \\
\hline
  prolonged and restricted variables & conserved variables: $D$, $S_i$, $\tau$ & $D$,
  $\tilde u_i(=S_i/D)$, $h$ \\
\hline
  interpolation for the prolongation & & Lagrangian: $5^{\rm th}$
    order in space (reduced to $1^{\rm st}$ order in \\ of the
    hydrodynamical variables & ENO: $3^{\rm rd}$ order in space,
    $2^{\rm nd}$ order in time & case of failure), $2^{\rm nd}$ order
    in time \\ & & (reduced to $1^{\rm st}$ order at extrema)\\
\hline
buffer zones & 12 & 6 \\
\hline
  overlapping same-level grids are & evolved as a single grid & evolved independently
  (but using the average of the values  \\ && of the two grids at overlapping points)\\
\end{tabular}
\end{ruledtabular}
\label{table:amr-implem}
\end{table*}

\subsubsection{Equations of state}
\label{sec:EoS}
In this work we present results obtained with two EoSs: a simple
``$\Gamma$-law'' or ``ideal-fluid'' EoS and a piecewise-polytropic
EoS~\cite{Read:2009a}. For the ideal-fluid EoS, the pressure is given
as
\begin{equation}
\label{id fluid}
p = (\Gamma-1) \rho\, \varep \ ,
\end{equation}
where $\Gamma$ is the adiabatic index. When using the ideal-fluid
EoS (\ref{id fluid}), nonisentropic changes can take place in the
fluid and, in particular, shocks (which are always present in the
mergers and which may play important roles) are allowed to transfer
kinetic energy to internal energy. On the other hand, a carefully
chosen piecewise-polytropic EoS may mimic more
closely a realistic EoS. The parametrised EoS we consider consists of
two polytropes interfacing at a density $\rho_0$. The relations
between the hydrodynamical quantities are ($i=0,1$)~\cite{Read:2009a}
\begin{eqnarray}
p &=& K_i \rho^{\Gamma_i}\ , \\
\varep &=& (1+a_i)\rho+\frac{K_i}{\Gamma_i-1}\rho^{\Gamma_i}\ ,
\label{varep0}
\end{eqnarray}
where $K_i$ are the polytropic coefficients, and $\Gamma_i$ are the polytropic exponents in the different
intervals of rest-mass density. Furthermore, the constants $a_i$, which guarantee continuity, are
\begin{eqnarray}
a_0 &=& 0\ , \\ a_1 &=& \frac{\varep(\rho_0)}{\rho_0}
-1-\frac{K_1}{\Gamma_1-1}\rho_0^{\Gamma_1-1}\ .
\end{eqnarray}
In our simulations we used the parameters of model B
of~\cite{Read:2009b}, namely:
\begin{equation}
\rho_0= 1.630497500125504\times 10^{14} \ \rm{g\ cm}^{-3}\ ,
\end{equation}
\begin{eqnarray}
0<\rho<\rho_0:\quad \Gamma_0&=&1.35692395\ , \nonumber \\
K_0&=&0.35938266\times 10^{14}\ \rm{cgs\ units},\nonumber \\
\rho>\rho_0:\quad \Gamma_1&=&3.0\ , \nonumber \\
K_1&=&  0.15982116\times 10^{-9}\ \rm{cgs\ units},\nonumber  \\
a_1&=&0.01088158737430845\ .\nonumber 
\end{eqnarray}

In the presence of shock heating, part of the kinetic energy is converted
into thermal energy. To model this property, the original
piecewise-polytropic EoS is modified by adding a thermal contribution to the pressure
\begin{equation} 
P_{\rm th}=(\Gamma_{\rm th}-1)\rho (\varepsilon-\varepsilon_0),
\end{equation}
where $\Gamma_{\rm th}$ is the adiabatic index for this 
correction and $\varepsilon_0$ is given by Eq.~(\ref{varep0}). 
In the absence of shocks, $\varepsilon$ is equal to $\varepsilon_0$ 
and thus $P_{\rm th}=0$. In the simulations of this work {\tt SACRA} 
has adopted $\Gamma_{\rm th}=\Gamma_0$ while in {\tt Whisky} the thermal 
correction was not applied ($P_{\rm th}=0$ ). As the figures of this work show, 
at least in the inspiral phase the difference in the adopted EoS does not have an influence.

\subsection{Adaptive Mesh Refinement}  
\label{sec:NumericalSpecifications}

There are similarities and differences in the implementation of the adaptive mesh refinement (AMR) in
the two codes. In the following we spell them out in detail.

Both codes employ a vertex-centered Berger--Oliger~\cite{Berger84}
mesh-refinement scheme adopting nested grids with a $2:1$ refinement
factor for successive grid levels. In the simulations made for the
present work both codes used a set of coarser fixed grids and finer
moving grids, centered around each star. {\tt Whisky} makes use of the
\texttt{Carpet} mesh-refinement driver~\cite{Schnetter-etal-03b}. The
higher-resolution moving grids are centered around the local maximum
in the rest-mass density $\rho$ of each star. In {\tt SACRA}, instead, the
grids are centered around the local maximum of the conserved variable
$D$.

Both codes employ centered $4^{\rm th}$-order finite-differencing in space
for evaluating spatial derivatives of the geometric quantities, except for the shift advection terms
that are calculated with upwinding derivatives to improve accuracy. For
the time integration, the $4^{\rm th}$-order Runge-Kutta scheme is adopted.
To evolve quantities near the refinement boundaries of a refined grid,
both codes introduce buffer zones, where the variables are computed
(``prolonged'' and ``restricted'') in a special way and not with the
time-update scheme used for all other non-refinement-boundary points.

In {\tt Whisky}/{\tt Carpet}, we use $12$ buffer points, $3$ for each
substep of the adopted time-integration scheme. The values of the
needed quantities at the buffer points are computed from the coarser
grid through interpolation as follows: For the spacetime variables,
$5^{\rm th}$-order Lagrangian interpolation in space and $2^{\rm nd}$-order
Lagrangian interpolation in time are used; For the hydrodynamical
variables, $3^{\rm rd}$-order ENO~\cite{Harten87} interpolation in space and
$2^{\rm nd}$-order ENO interpolation in time are used.  The prolonged and
restricted variables are the conserved evolved ones: $D$, $S_i$, and
$\tau$. The interpolation is done whenever the first Runge-Kutta 
time integration is being carried out. 

In {\tt SACRA} both prolongation and restriction are carried out on
$D$, $\tilde u_i(=S_i/D)$, and $h$. Following~\cite{Brugmann:2007zj},
$6$ buffer points are introduced. The quantities at the buffer zones
are provided from the corresponding coarser domain by the following
procedure. For space interpolation, $5^{\rm th}$-order centered Lagrangian
interpolation in space is carried out using the nearby $6$ points of
the coarser grid. This is done both for spacetime and hydrodynamics
variables. For the latter, this interpolation scheme could fail, in
particular in the vicinity of the surface of the stars, where $D$ is small
and varies steeply. The reason for this possible failure is that the
interpolation may give a negative, and so unphysical, value of $D$ or
$h-1$. If the $5^{\rm th}$-order Lagrange interpolation produces $D <
D_{\rm min}$ or $h <1$, $1^{\rm st}$-order (i.e., linear) interpolation is
adopted. $D_{\rm min}$ is chosen to be $D_{\rm max}/10^9$, where
$D_{\rm max}$ is the initial maximum value of $D$. Linear
interpolation cannot be used in general for all points because it is
too dissipative. As in {\tt Whisky} the interpolation is also done whenever the first
Runge-Kutta time integration is being carried out. 

For the update of the buffer zones {\tt SACRA} implements, instead,
the following: i) For the inner three buffer points all the quantities
are evolved using the $4^{\rm th}$-order finite-differencing scheme. Since
there is a sufficient number of buffer points to solve the evolution
equations in the inner three buffer points, no interpolation is
necessary; ii) For the fourth buffer point, all the quantities are
evolved using a $4^{\rm th}$-order finite-differencing scheme with no
interpolation, except for the transport terms for the geometry such as
$\beta^k\pa_k \tilde \gamma_{ij}$, for which $2^{\rm nd}$-order finite
differencing is employed when $\beta^k$ has an unfavorable
sign; iii) For the two outer buffer points, $2^{\rm nd}$-order Lagrangian
interpolation in time of the coarser-grid quantities is carried
out. This time-integration procedure is applied to both spacetime and
hydrodynamical variables, but for the latter there is an additional
check.

The interpolated value at a finer-grid time step is obtained from the
values at the three time levels of the coarser grid, say, $n-1$, $n$,
and $n+1$ (note that $n$ does not denote the Runge-Kutta time step).
The interpolation is necessary for determining the values at a time
$t$ that satisfies $t^n < t < t^{n+1}$. Defining $Q$ as $D$, $\tilde
u_i$, or $h$, and $Q^n$ as the value of the variable $Q$ at time
$t^n$, {\tt SACRA} checks whether ($Q^{n+1}-Q^n)(Q^n-Q^{n-1}) < 0$ and
if so adopts $1^{\rm st}$-order interpolation, using only $Q^{n+1}$ and
$Q^n$. Namely, a limiter procedure is introduced. This robust
prescription provides numerical stability~\cite{Yamamoto2008}.

\begin{table*}[ht]
  \caption{Properties of the initial data: proper separation between
    the centers of the stars $d/M_{_{\rm ADM}}$; baryon mass $M_{b}$
    of each star in units of solar mass; total ADM mass $M_{_{\rm ADM}}$ in
    units of solar mass, as measured on the finite-difference grid; total ADM
    mass $\tilde{M}_{_{\rm ADM}}$ in units of solar mass, as provided by the
    Meudon initial data; angular momentum $J$, as measured on the
    finite-difference grid; angular momentum $\tilde{J}$, as provided
    by the Meudon initial data; initial orbital angular velocity
    $\Omega_0$; mean coordinate equatorial radius of each star $r_e$
    along the line connecting the two stars; maximum rest-mass density
    of a star $\rho_{\rm max}$. The columns for $M_{\rm ADM}$ and $J$ contain the value for {\tt
      Whisky} (left) and the one for {\tt SACRA} (right). Note that the values of $M_{\rm ADM}$ 
    and $J$ are computed through a volume
    integral in {\tt Whisky}, while in  {\tt SACRA} they are computed through the extrapolation to $r \rightarrow
    \infty$ of the ADM masses and angular momenta calculated as surface integrals at finite radii $r$.}
\begin{ruledtabular}
\begin{tabular}{l|ccccccccccc}
EoS for the model &
\multicolumn{1}{c}{$d/M_{_{\rm ADM}}$} &
\multicolumn{1}{c}{$M_{b}$} &
\multicolumn{1}{c}{$M_{_{\rm ADM}}$} &
\multicolumn{1}{c}{$\tilde{M}_{_{\rm ADM}}$} &
\multicolumn{1}{c}{$J$} &
\multicolumn{1}{c}{$\tilde{J}$} &
\multicolumn{1}{c}{$\Omega_0$} &
\multicolumn{1}{c}{$r_e$} &
\multicolumn{1}{c}{$\rho_{\rm max}$} \\
~ &
\multicolumn{1}{c}{$~$} &
\multicolumn{1}{c}{$(M_{\odot})$} &
\multicolumn{1}{c}{$(M_{\odot})$} &
\multicolumn{1}{c}{$(M_{\odot})$} &
\multicolumn{1}{c}{$(\times10^{49}{\rm g\, cm}^2{\rm /s})$} &
\multicolumn{1}{c}{$(\times10^{49}{\rm g}\, {\rm cm}^2{\rm /s})$} &
\multicolumn{1}{c}{$({\rm rad/ms})$} &
\multicolumn{1}{c}{$({\rm km})$} &
\multicolumn{1}{c}{$({\rm g/cm}^3)$} \\
\hline
Ideal fluid ($\Gamma=2$) & $12.6$ & $1.779$ & $3.251,3.256$ & $3.233$ 
& $8.921,8.930$ &
$8.922$ & $1.906$ & $12.23$ & $7.58\times10^{14}$\\
Piecewise polytropic & $15.4$ & $1.502$ & $2.676,2.680$ & $2.668$ 
& $6.492,6.506$ & 
$6.491$ & $1.664$ & $\ \ 8.48$ & $9.77\times10^{14}$\\
\end{tabular}
\end{ruledtabular}
\vskip -0.25cm
\label{table:ID}
\end{table*}

The two domains in the finer levels often overlap. In such cases, the
values of all quantities should agree with each other, but, since in {\tt SACRA} the
evolution equations for the two domains are solved independently, the
values do not always agree exactly. Let us denote with $Q_1$ and
$Q_2$ the values on the two domains of an individual refinement
level. In order to guarantee that they are the same, in {\tt SACRA}
the average of the two values is used: $Q_1= Q_2 \rightarrow (Q_1 +
Q_2)/2$. When a buffer point of one of the two domains overlaps with a
point in the main region of the other domain, the values at the point
of the main region are copied to those at the buffer point. When two
buffer zones overlap at some points, the simple averaging described
above is again used.

In {\tt Whisky}, when domains of the same refinement level would overlap, the whole level is
automatically resplit in (smaller and more numerous) nonoverlapping domains, so in practice they
continue to be evolved as a single grid, without requiring averaging. For more details on the {\tt
  Carpet} code see~\cite{Schnetter-etal-03b}.

For both codes, at the outer boundaries of the coarsest refinement
level, an outgoing boundary condition is imposed for all the geometric
variables. The outgoing boundary condition is the same as that
suggested by Shibata and Nakamura \cite{Shibata95}. Flat boundary
conditions are applied to the matter variables.

Both codes can add artificial dissipation to the source terms of the
Einstein equations. In particular, for the schemes presented in this
work, they could use $5^{\rm th}$-order Kreiss-Oliger-type
dissipation~\cite{Kreiss73} as $Q_l \rightarrow Q_l - \sigma
h_l^6 Q^{(6)}_l$ where $Q_l$ is a quantity in the $l$-th level, $h_l$
is the spacing of the $l-$th level, $Q^{(6)}_l$ is the sum of the
sixth derivatives along the $x$, $y$, and $z$ axis directions, and
$\sigma$ is a constant of order $0.1$.  The results of the present
work were obtained without artificial dissipation for {\tt SACRA} and
with artificial dissipation for {\tt Whisky}.

Standard {\tt SACRA} simulations for NS-NS binaries are performed with
$7$ or $8$ refinement levels, in particular $3$ or $4$ coarser levels
composed of one domain and $4$ finer levels composed of two
domains. The time step for each refinement level, $dt_l$, is
determined as follows:
\beqn
dt_l=\left\{
\begin{array}{ll}
h_{2} /2 & {\rm for}~0 \leq l \leq 2 \\ 
h_l /2 & {\rm for}~2 < l \leq L-1.  \\ 
\end{array}
\right. \label{eqdt}
\eeqn
Namely, the Courant number (expressed in terms of the speed of light)
is $1/2$ for the finer refinement levels with $l \geq 2$, whereas
for the coarser levels, it is smaller than $1/2$. The reason why a
smaller Courant number is chosen for the coarser levels is that with a
Courant number as high as $1/2$, numerical instabilities occur near
the outer boundary. This is an inherent problem of the adopted $\tilde{\Gamma}$-driver gauge
condition~\cite{Schnetter:2010} and it does not appear in the {\tt Whisky} simulations of the
present work only because the resolution in the coarsest grids is still high enough.

In standard {\tt Whisky} simulations for binary systems $6$ refinement
levels are used, the two finest of which move following the stars. In
addition to the moving grids, a set of refined but fixed grids is set
up at the center of the computational domain so as to capture the
details of the Kelvin-Helmholtz instability (see~\cite{Baiotti08}). 
The Courant factor is $0.35$ for all levels.

In the {\tt Whisky} simulations for the present work, a reflection symmetry 
condition across the $z=0$ plane and a
$\pi$-symmetry condition\footnote{Stated differently, we evolve only
the region $\{x\geq 0,\,z\geq 0\}$ applying a $180^{\circ}$-rotational-symmetry
boundary condition across the plane at $x=0$.} across the $x=0$ plane
are used, while {\tt SACRA} adopts only the reflection symmetry across the $z$ plane.

The differences in the implementation of AMR between {\tt SACRA} and
{\tt Whisky} are summarized in Table~\ref{table:amr-implem}.

\begin{table*}[t]
  \caption{Properties of the initial grids: number of refinement
    levels (including the coarsest grid); number of finer levels 
    that are moved to follow the stars; spacing of the finest level;
    length of the side of the finest level; spacing of the coarsest
    level; outer-boundary location. All lengths are expressed in
    km. HR, MR, and LR denote the high, medium, and low resolutions, respectively. 
    For {\tt Whisky} the two resolutions are in a ratio of 5/4,
    while for {\tt SACRA} the ratio between LR and MR is 50/42 and the ratio between MR and HR is 1.16.}
\begin{ruledtabular}
\begin{tabular}{l|c|c|c|c|c|c}
Model  & ${\rm n^o}$ of levels & ${\rm n^o}$ of moving levels & finest spacing 
& extent of finest grid & coarsest spacing & outer-boundary
 location \\
\hline
{\tt Whisky} ideal fluid   & $6$ & 2 & $0.1773$ & $44.33$ & $\ \ 5.67$ & $380$ \\
{\tt Whisky} piecewise, HR & $6$ & 2 & $0.1773$ & $44.33$ & $\ \ 5.67$ & $760$ \\
{\tt Whisky} piecewise, LR & $6$ & 2 & $0.2216$ & $44.33$ & $\ \ 7.09$ & $760$ \\
{\tt SACRA} ideal fluid   & $8$ & 4 & $0.1387$ & $\ \ \ \  6.656$ & $17.75$ & $852$ \\
{\tt SACRA} piecewise, HR & $7$ & 4 & $0.1746$ & $\ \ \ \  9.428$ & $11.17$ & $603$ \\
{\tt SACRA} piecewise, MR & $7$ & 4 & $0.2025$ & $10.13$ & $12.96$ & $648$ \\
{\tt SACRA} piecewise, LR & $7$ & 4 & $0.2411$ & $10.13$ & $12.96$ & $648$ \\
\end{tabular}
\end{ruledtabular}
\vskip -0.25cm
\label{table:grids}
\end{table*}

\subsection{Initial data}
\label{sec:initial_data}

The initial configurations for our relativistic-star binary
simulations are produced using the multidomain spectral-method code,
{\tt LORENE}, which was originally written by the group working at the
Observatoire de Paris-Meudon~\cite{Gourgoulhon01,Taniguchi02b} and
which is publicly available~\cite{lorene}. Specific routines are used
to transform the solution from spherical coordinates to a Cartesian
grid of the desired dimensions and shape.

These initial data, which we refer to also as the {\it ``Meudon
data''}, are obtained under the assumptions of quasiequilibrium and
of conformally-flat spatial metric. The initial data used in the
simulations shown here were produced with the additional assumption of
irrotationality of the fluid flow, {\it i.e.}  the condition in which
the spins of the stars and the orbital motion are not locked; instead,
they are defined so as to have vanishing vorticity. Initial data
obtained with the alternative assumption of rigid rotation were not
used because, differently from what happens for binaries consisting of
ordinary stars, relativistic-star binaries are not thought to achieve
synchronization (or corotation) in the timescale of the
coalescence~\cite{Bildsten92}.

The initial models for the binaries have been chosen so as to allow
significant possibilities of comparison between the codes and at the
same time to limit the required computational time. In particular, after performing several orbits
and merging, prompt collapse to a Kerr black hole occurs. As said above, we chose two EoSs, the
ideal-fluid EoS~\eqref{id fluid}\footnote{The initial data for the simulations adopting the
  ideal-fluid EoS are set up as a simple polytropic EoS with polytropic constant $K = 123.6$ 
  (in units of $c=G=M_{\odot}=1$).} 
and a piecewise-polytropic EoS~\eqref{varep0}. For the latter, the
initial data have been kindly provided by K. Taniguchi. Note that the model with
the ideal-fluid EoS has been often used in previous work
({\it e.g.}~\cite{Baiotti:2009gk,Yamamoto2008}).

Some of the physical quantities of the initial configurations are reported in
Table~\ref{table:ID}. In brief, they are equal-mass configurations with an initial proper distance
between stellar centers of about $60$ km (initial orbital frequency $0.303$ kHz and $0.265$ kHz,
respectively for the ideal-fluid model and for the piecewise-polytropic model). The chosen rest masses
of $M_0^{\rm IF}=1.779 M_\odot$ and $M_0^{\rm PP}=1.502 M_\odot$, respectively for the two models, lead - as desired -
to prompt collapse to black hole.

\subsection{Specific grid setup for the reported simulations}

For the higher-resolution run with {\tt Whisky}, the spacing of the
finest of the six grid levels is $h=0.120\,M_{\odot}\simeq 0.1773\,\km$
and the spacing in the wave zone (the coarsest grid) is
$h=3.84\,M_{\odot}\simeq 5.67\,\km$. For the lower-resolution run the
spacing is $h=0.150\,M_{\odot}\simeq 0.2216\,\km$ on the finest grid
and $h=4.80\,M_{\odot}\simeq 7.09\,\km$ on the coarsest grid. The
finest grid always covers the whole stars. For the simulations with
the ideal-fluid model the outer boundary is located at about $380\km$
while in the case of the piecewise-polytropic model, for both resolutions,
the outer boundary is at about $760\ \km$. Except for the outer
boundary location and the grid spacing, the AMR grid structure was the
same for all the runs.

For the runs with {\tt SACRA}, for the ideal-fluid model, the grid
structure is essentially the same as in~\cite{Yamamoto2008}; the
finest of the eight grid levels has $h=0.0938\,M_{\odot}\simeq
0.1387\,\km$. For the simulations with the piecewise polytrope, the
computational domain is composed of seven grid levels with the finest grid
resolution being $h=0.1182\,M_{\odot} \simeq 0.1746\,\km$ at the high resolution,
$h=0.1370\,M_{\odot} \simeq 0.2025\,\km$ at the medium resolution, and $h=0.1631\,M_{\odot}
\simeq 0.2411\,\km$ at the lower resolution. The
resolution in the wave zone (for the coarsest grid level) is $h \simeq
11.17\,\km$ for the high-resolution run and $h \simeq 12.96\,\km$ for the others. 
The boundary of the finest grid is at $60\%$ of the
stellar radius (along a coordinate axis, at $t=0$) while the second
finest grid covers all the stars for the run with the ideal-fluid EoS,
whereas for the run with the piecewise-polytropic EoS, the finest grid
covers the stellar radius completely (the boundary of the finest grid
is at $115\%$ of the stellar radius).  The outer boundary is at about $852\ \km$ 
for the simulations performed with the ideal-fluid EoS and
at about$603\ \km $ or $648\ \km $ for those performed with the piecewise-polytropic
EoS, for the high-resolution run and the other runs respectively.

As already noted in Sec.~\ref{sec:NumericalSpecifications}, another
difference between the grid setups of the two codes is the adopted
symmetry. Both codes compute only the $z \geq 0$ portion of the
$\{x,y,z\}$ Cartesian coordinate numerical domain, but, while {\tt
SACRA} calculates all the $z \geq 0$ portion, {\tt Whisky} calculates
only the $x\geq 0$ part of the remaining domain, taking advantage of
the $180^{\circ}$ degree rotation symmetry characterising equal-mass binaries.

The properties of the grids adopted in the simulations with the two
codes are summarised in Table~\ref{table:grids}.

For the setup of the piecewise-polytrope high-resolution run, {\tt Whisky}, 
which heavily exploits large parallel facilities, uses approximately $22\times 10^6$
grid points and the total required memory for the high-resolution run is about 640 GBytes.
{\tt SACRA}, instead, which has been specifically developed for
being able to perform production simulations even on a laptop computer, uses about $7\times 10^6$ grid points and
about 11.6 GBytes of memory. 
For {\tt Whisky}, the total CPU time for the high-resolution piecewise-polytrope run was
about 450 CPU hours on 320 processors of the Ranger cluster (at the Texas Advanced Computing Center; 
the processors are AMD Opteron Quad-Core 64-bit, with clock frequency 2.3 GHz)
and for {\tt SACRA} it was about 2000 CPU hours on a
Quad-Core machine of Core-i7X processors with clock frequency 3.33 GHz.

\begin{figure}[b]
\begin{center}
   \includegraphics[width=0.45\textwidth]{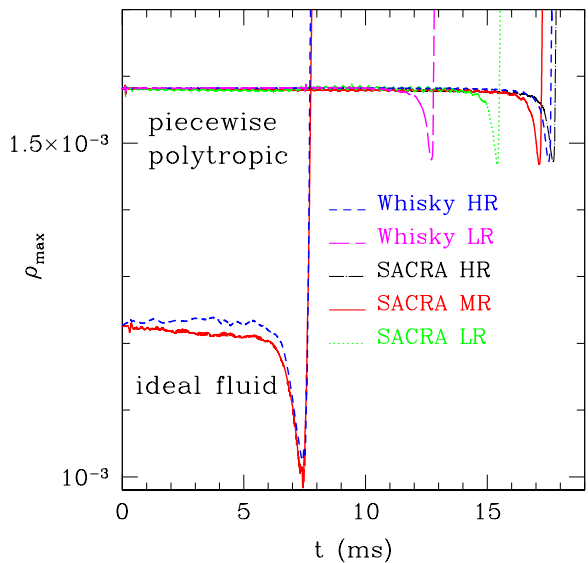}
\end{center}
\vskip -1cm
\caption{(Colour online) 
Comparison of the time evolution of the maximum of the rest-mass density for the two models (with
different EoSs) described in Sec.~\ref{sec:EoS}. For ease of interpretation, we remind the reader that in our 
units $\rho=1\times 10^{-3}$ corresponds approximately to $6.18 \times 10^{14}~{\rm g/cm^3}$. 
\label{fig:rho}}
\end{figure}

\begin{figure*}[t]
\begin{center}
   \includegraphics[width=0.45\textwidth]{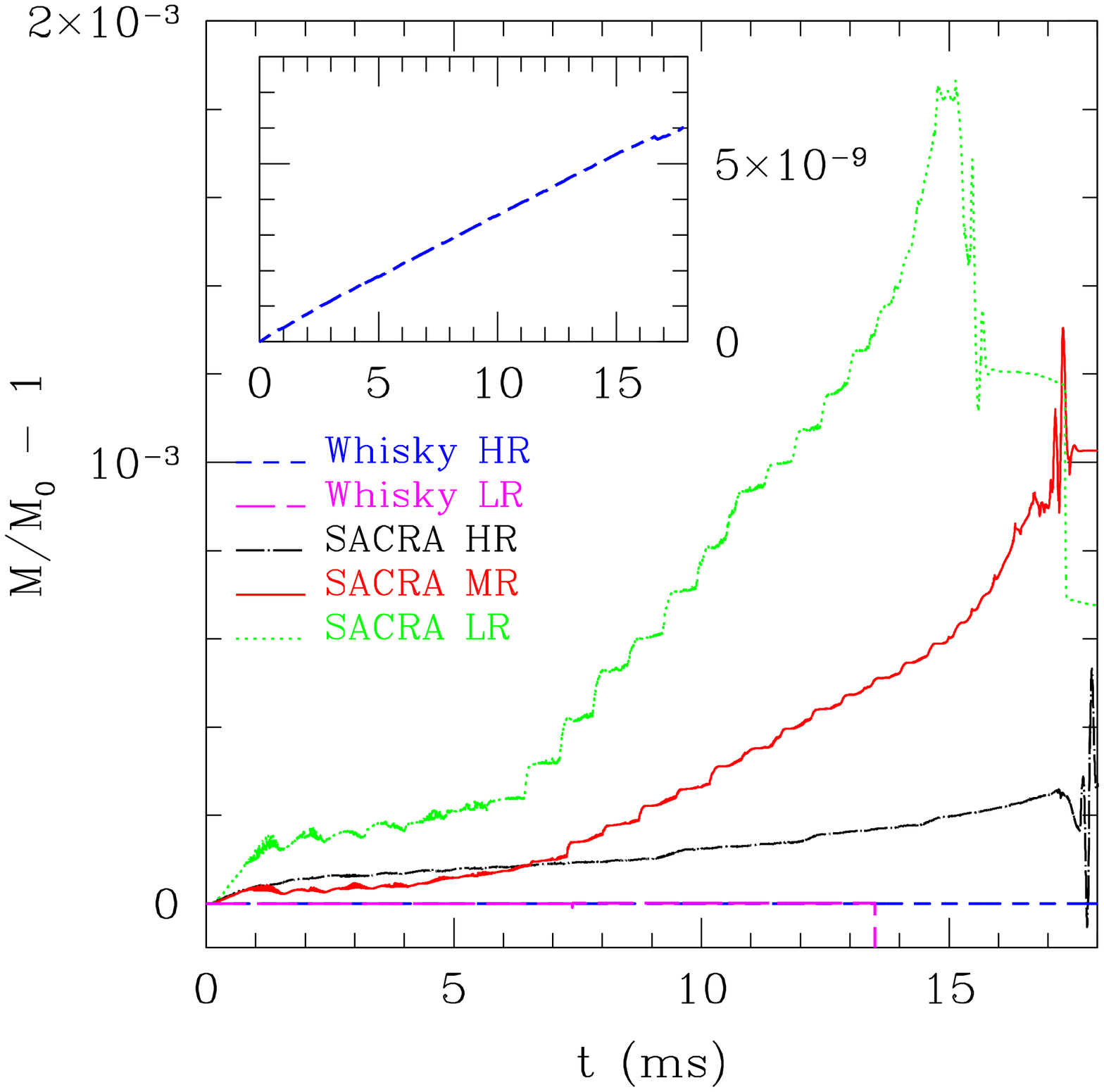}
   \hspace{1 cm}
   \includegraphics[width=0.45\textwidth]{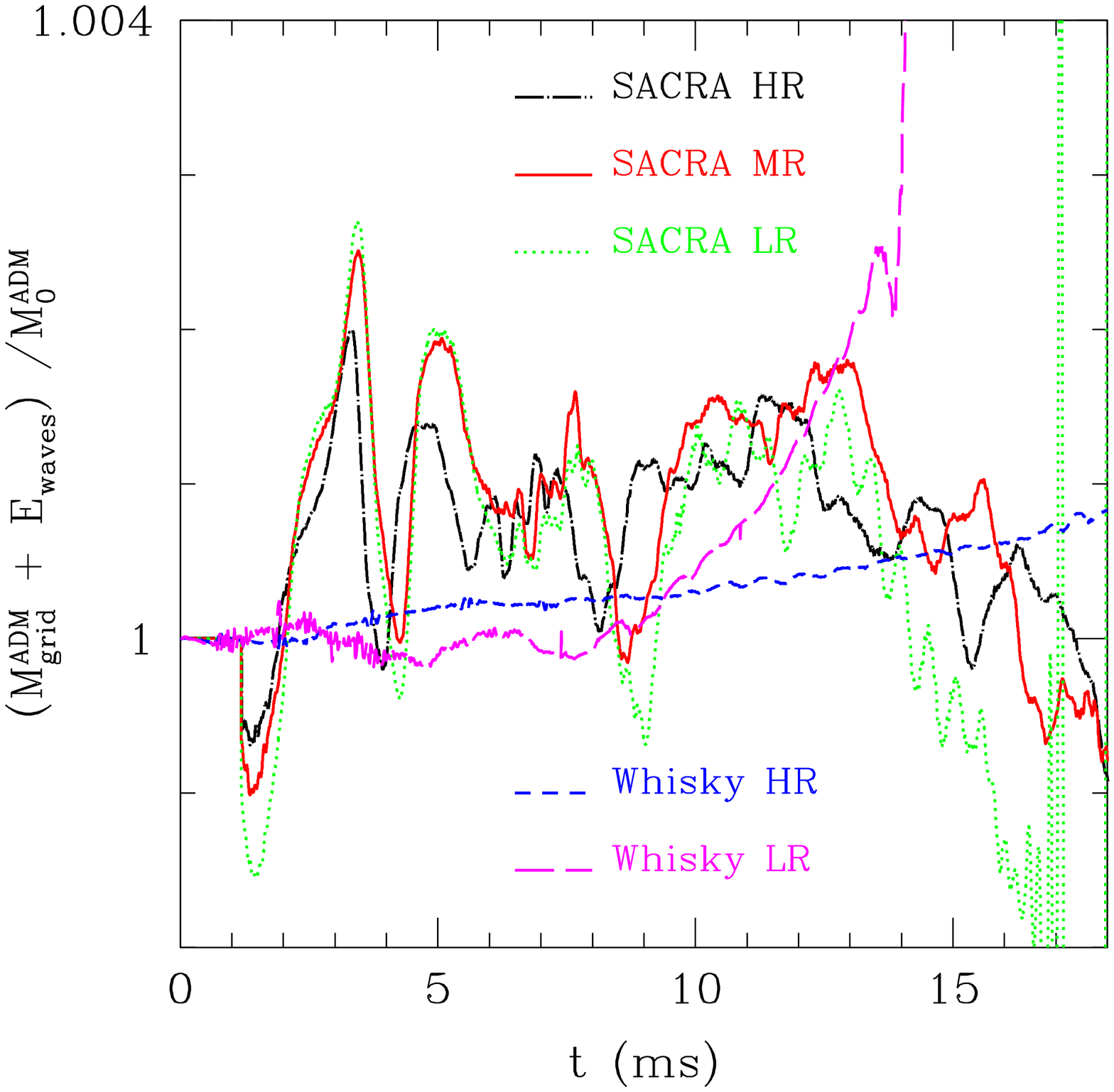}
\end{center}
\vskip -1cm
\caption{(Colour online) Left: Comparison of the time evolution of the rest mass (normalised to the
  initial value). The inset is a magnification of the higher-resolution {\tt Whisky} curve, in the
  form $M/M_0-1$. These data refer to the piecewise-polytropic EoS. As explained in the text,
  the larger variations in the SACRA data are due to the choice of grid structure. Right: Comparison
  of the time evolution of the sum (normalised to the initial value) of the ADM mass measured on the
  numerical grid and the energy carried away from the grid by gravitational waves. This quantity
  should be conserved. These data refer to the piecewise-polytropic EoS. Note that the data for the
  low resolution of {\tt Whisky} are not reliable after the formation of the AH ($t\simeq 13.4$ ms
  for this simulation), because the volume integral with which the ADM mass is computed contained
  also the points inside the horizon. See text for more details. \label{fig:masses}}
\end{figure*}

\section{Comparison of the results}

As also described in~\cite{Baiotti08,Yamamoto2008}, the chosen initial
data for the present study are such that the stars orbit about 3
times and 7 times, respectively for the two models with different EoSs, before merging.
As can be seen in Fig.~\ref{fig:rho}, the rest-mass density at the
stellar centres remains approximately constant for the first $6$ ms
(in the case of the simulation with the ideal-fluid EoS) or $15$ ms
(for the piecewise-polytropic case) and then decreases, indicating an
expansion of the stars due to the tidal force, just before the
merger. As expected from the (high) mass of the chosen models, the
merged object then immediately collapses to a black hole and the AH is
measured for the first time at about $8$ and $18$ milliseconds,
respectively for the two models with different EoSs (\cf the highest 
resolutions). The mass $M_{\rm BH}$ and the angular-momentum parameter 
$a \equiv J_{\rm BH}/(M_{\rm BH})^2$ of the resulting black hole are
measured by both codes. The values after the ringdown for the piecewise-polytropic EoS 
are $M^{\tt Whisky}_{\rm BH}=2.633M_\odot$, $M^{\tt SACRA}_{\rm  BH}=2.637M_\odot$ (a 
relative difference of $0.15\%$), and $a^{\tt Whisky}=0.79$, $a^{\tt SACRA}=0.80$ (a relative
difference of $1.2\%$). 
For the ideal-fluid EoS the values of the black hole are $M^{\tt Whisky}_{\rm BH}=3.22M_\odot$, 
$M^{\tt SACRA}_{\rm  BH}=3.21M_\odot$, and $a=0.84$ 
for both codes. \\

Having briefly summarised the dynamics of the system, we present now
first a comparison between some quantities produced in evolutions
performed with {\tt SACRA} and with {\tt Whisky}, each in what is
thought to be a good configuration in terms of accuracy, violation of
the ADM constraints, and cleanness of gravitational waves.
Furthermore, for the piecewise-polytropic EoS we present for each code results obtained at 
two or three resolutions.

\begin{figure}[t]
\begin{center}
   \includegraphics[width=0.45\textwidth]{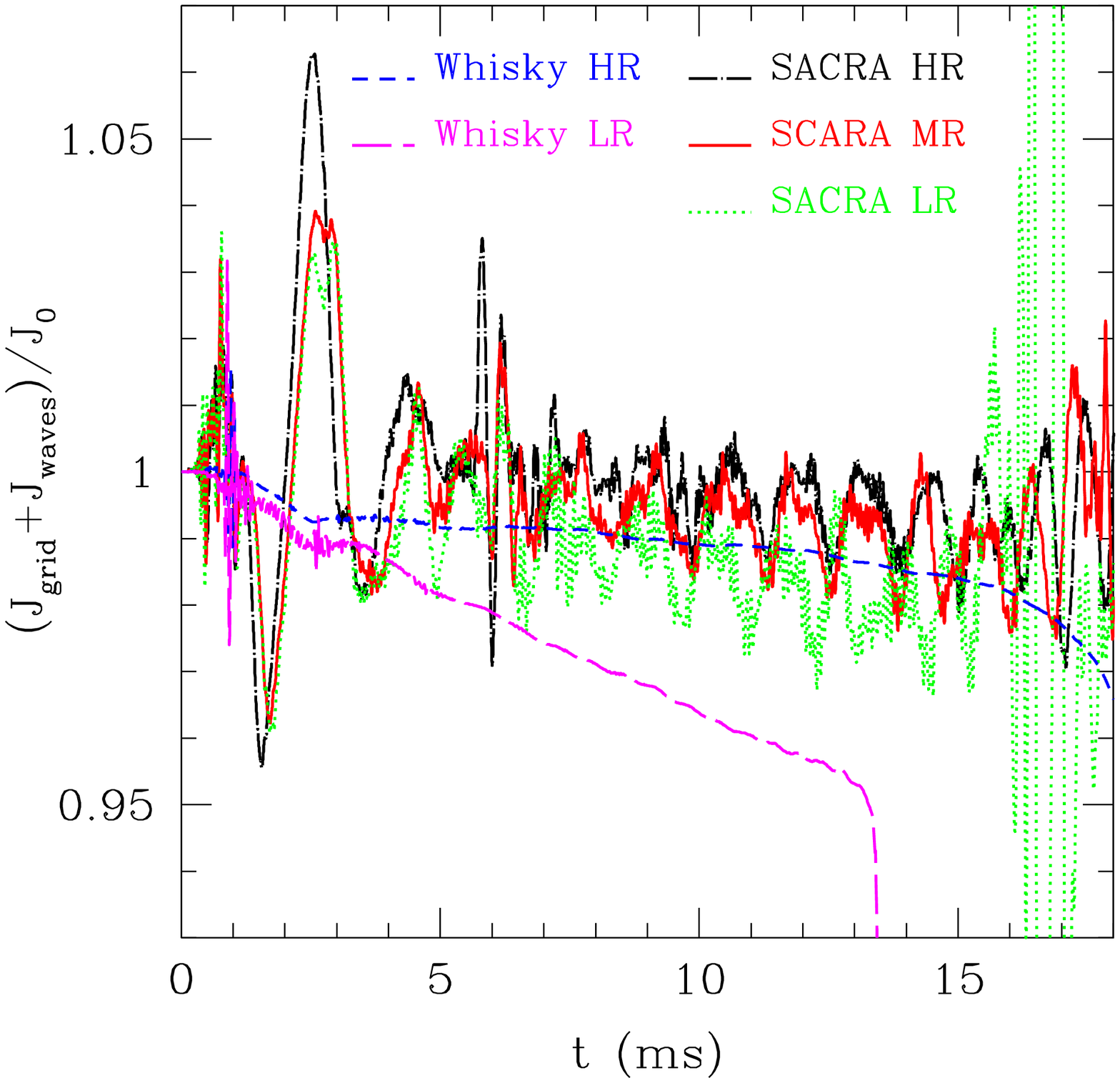}
\end{center}
\vskip -1cm
\caption{(Colour online) 
  Comparison of the time evolution of the angular momentum, computed as
  the sum of the angular momentum measured on the numerical grid and the
  angular momentum carried away from the grid by gravitational
  waves. These data refer to the piecewise-polytropic
  EoS. As already noted for Fig.~\ref{fig:masses}, also here the data for the
  low resolution of {\tt Whisky} are not reliable after the formation of the AH ($t\simeq 13.4$ ms
  for this simulation), because the volume integral with which the angular momentum is computed excludes the
  contribution of the black hole. See text for more details. \label{fig:am}}
\end{figure}

\begin{figure*}[t]
\begin{center}
   \includegraphics[width=0.45\textwidth]{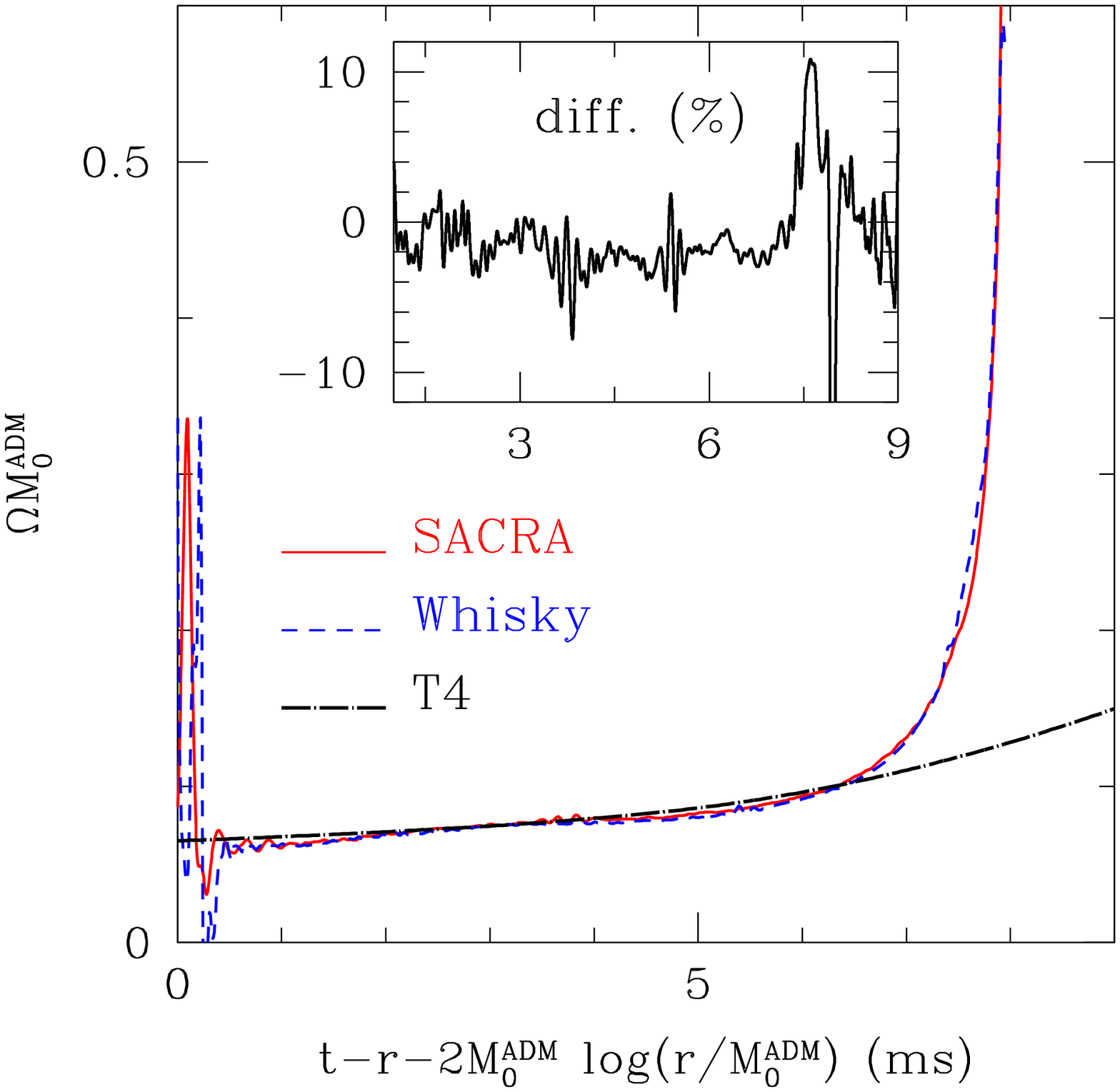}
\hspace{1 cm}
   \includegraphics[width=0.45\textwidth]{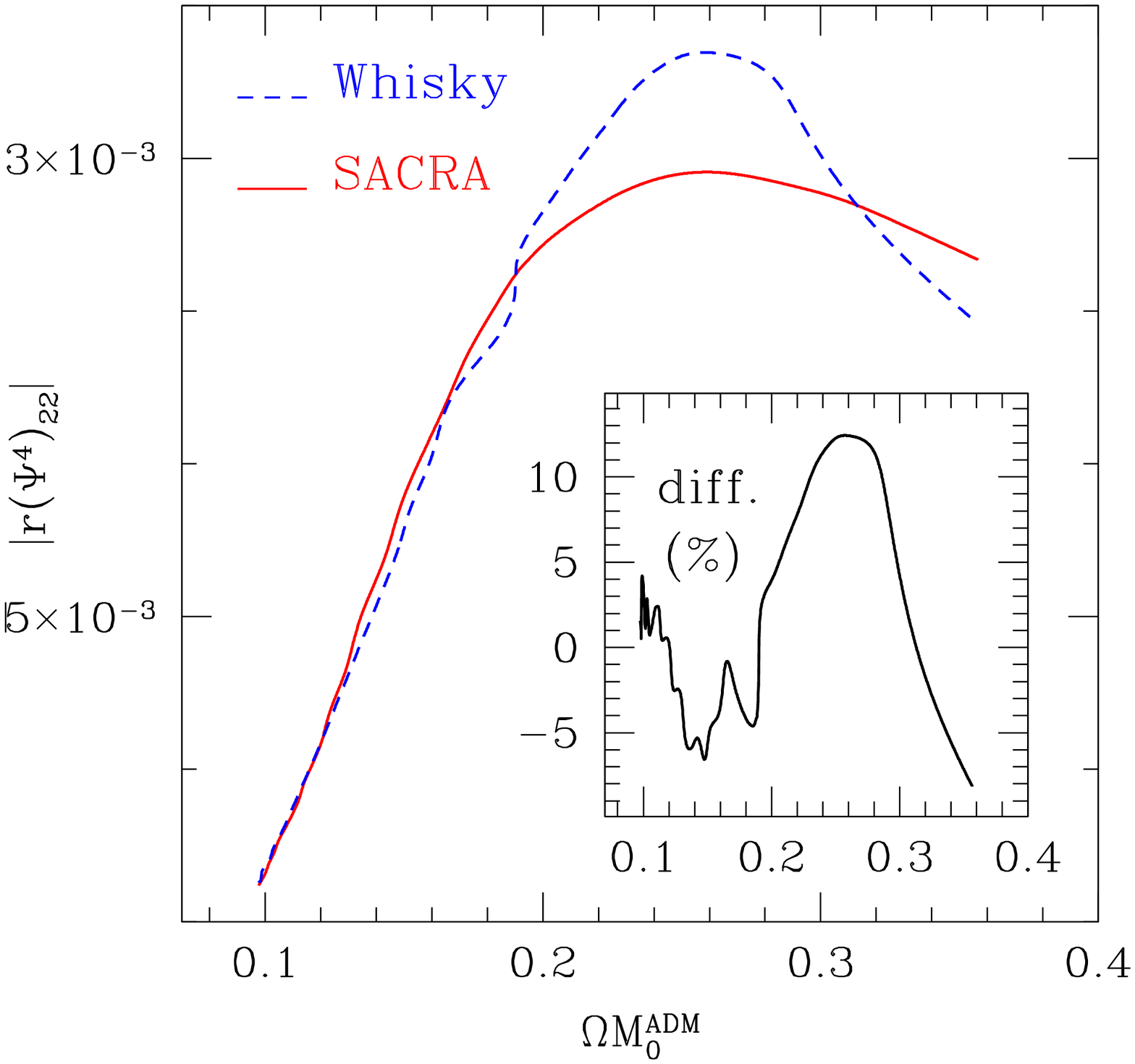}
\end{center}
\vskip -1cm
\caption{(Colour online) Left: Comparison of the time evolution of the 
 orbital frequency, computed from $\psi_4$. These data refer to the
  ideal-fluid EoS. The inset shows the percent difference of the two
  curves. 
  The curve labeled T4 is the Taylor-T4 post-Newtonian approximation~\cite{Baker-etal:2007b,Boyle:2007ft,Baiotti:2010b}.
  Right: Comparison of the amplitude of the wave as a function
  of the frequency $M\Omega$. The inset shows the percent difference
  between the two curves. These data refer to the ideal-fluid EoS.}
  \label{fig:freq}
\end{figure*}

From Fig.~\ref{fig:rho} one can see immediately that the time of the
merger depends considerably on the grid resolution, for both codes,
but in a stronger fashion for {\tt Whisky}. As is well known, the
conservation of the angular momentum in numerical simulations of
binary compact objects is a delicate issue, which can have very
visible effects like the ones in Fig.~\ref{fig:rho}. Even if the
merger and post-merger dynamics may not be sensible to the exact timing of
the inspiral, the phase of gravitational waves is affected and
so this effect must be carefully taken into account when producing
templates for gravitational-wave data analysis. For example,
\cite{Yamamoto2008} attempted to do so by estimating, given a specific
initial-data configuration, the 'real' merger time at infinite
resolution through an extrapolation based on the results of
simulations of the same model at different resolutions.  Anyway, we
are here interested in the comparison of the codes and note that, when
the differences due to the resolution are subtracted by time-shifting
the curves, the evolutions of the rest-mass density in the two codes
are very similar. As said, a proper analysis of the phase difference of the gravitational waves from
the various codes and resolutions will be reported in a future work~\cite{Baiotti:2010b}.

\begin{figure}[t]
\begin{center}
   \includegraphics[width=0.45\textwidth]{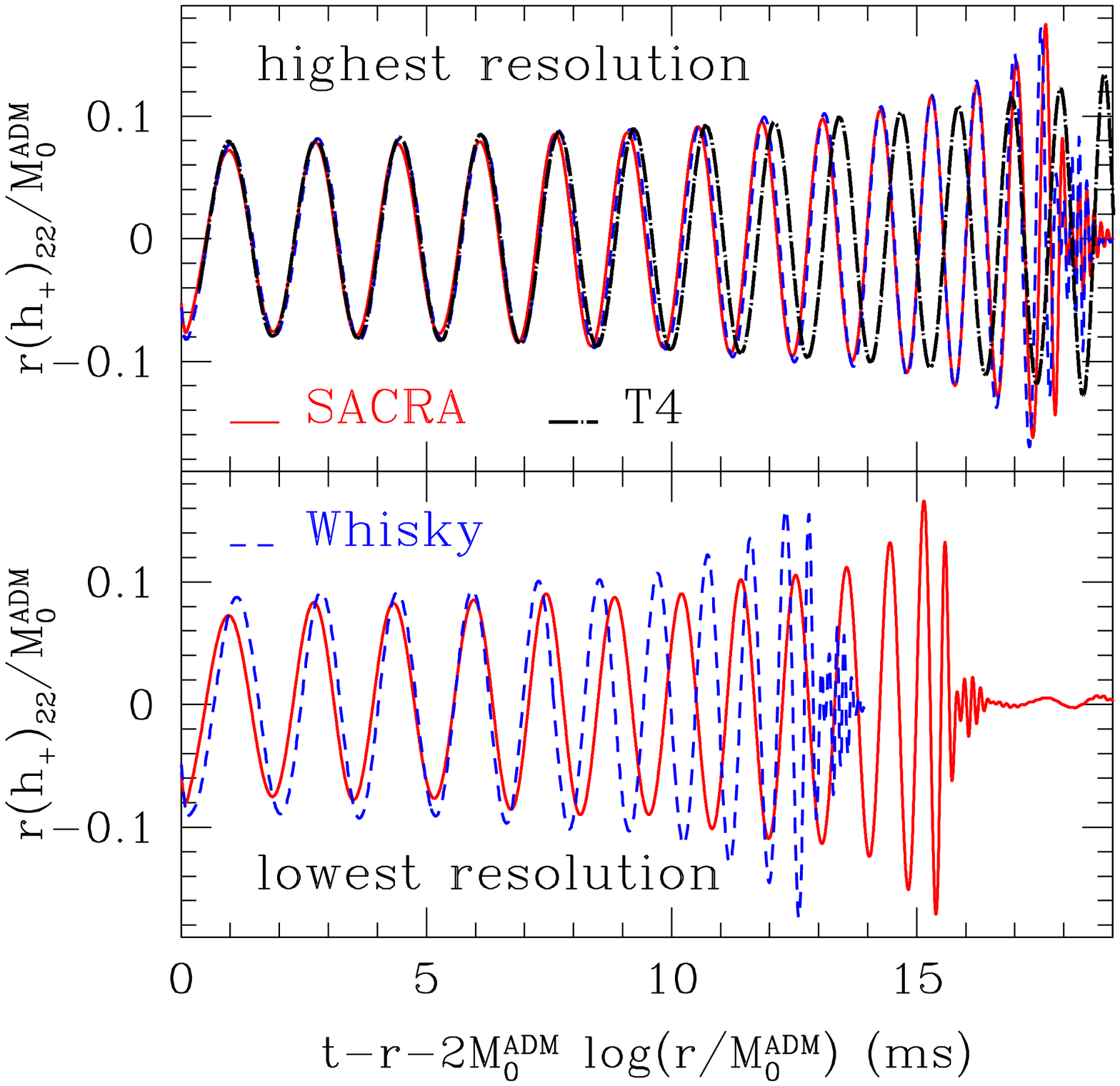}
\end{center}
\caption{(Colour online) Comparison of the waveform $(h_+)_{22}$. 
  These data refer to the piecewise-polytropic EoS. The upper panel
  refers to the ``higher resolution'' and the lower panel to the
  ``lower resolution'' (see text for details). As in Fig.~\ref{fig:freq}, 
  the curve labeled T4 is the Taylor-T4 post-Newtonian 
  approximation~\cite{Baker-etal:2007b,Boyle:2007ft,Baiotti:2010b}. } \label{fig:h+}
\end{figure}

We continue the discussion of the results in a more quantitative
manner by comparing the time evolution of the rest mass, which should
be a conserved quantity as no matter is seen leaving the numerical
domain through the outer boundary during the simulation. One can see
in the left panel of Fig.~\ref{fig:masses} that both codes conserve
the rest mass at very high accuracy, but in {\tt SACRA} the violation
is of the order of $10^{-3}$ while in {\tt Whisky} it is of the order
of $10^{-8}$. More in detail, the dot-dashed black line refers to the 
high-resolution {\tt SACRA} run, which of course shows an
improvement over the medium (continuous red line) and lower-resolution ones (dotted green
line). The convergence is achieved approximately at second order. 
The curves referring to {\tt Whisky} look constant on the main
panel, but in the subpanel one can notice the minute increase in the
rest mass even in the high-resolution results (short-dashed blue
line). The curve referring to the low resolution (long-dashed magenta)
drops at the time of AH formation because the matter inside the
horizon is not included in the computation of the rest mass.

The reason of the relatively worse conservation in {\tt SACRA} (as
said, the conservation is very good in absolute terms also for {\tt
SACRA}) is to be found in the presence of a refinement boundary very
close to the stellar surface. In the orbital phase, oscillations due to the tidal deformation of the
NS cause the matter to cross the finest refinement level and the small errors due to the interpolation in
the buffer zones are larger where the density is larger. Also in {\tt
Whisky}, if a refinement boundary is placed inside the
stars, the violation of the conservation of the rest mass is larger
($\sim10^{-4}$).

The right panel of Fig.~\ref{fig:masses}, which - as the left one - refers to the piecewise-polytropic 
EoS, shows then the conservation
of the energy, namely the sum of the ADM mass computed on the
numerical domain and of the energy carried by gravitational waves
outside the numerical domain. Such a quantity, normalised to its
initial value (the initial ADM mass) should be constant and the figure
shows the deviation of the results from constancy. The colours and
line types are the same as in the left panel. At the highest
resolutions, both {\tt Whisky} and {\tt SACRA} conserve this quantity
very well, at the order of 1 per 1000 during the inspiral and at
better than 1\% overall.

Some of the differences in the curves referring to the two codes (in
particular the 'smoothness') are due to the different way of computing
the ADM mass. {\tt Whisky} performs a volume integral with the formula
\begin{equation}
M_{_{\rm ADM, Vol}} = \int_V \de_i \Big [\alpha \sqrt{\gamma}
\gamma^{jk}\gamma^{li}(\de_k\gamma_{jl} - \de_l\gamma_{jk}) \Big ]
d^3x 
\end{equation}
and in the simulations of this work it does not exclude the points
inside the AH from the computation, so the values of the ADM mass
given by {\tt Whisky} after the appearance of the AH are affected by
gross errors. {\tt SACRA}, instead, uses a surface integral on a
spherical surface far from the central objects. This method gives
consistent results after the formation of the AH, but is more sensitive to
small metric oscillations in the vicinity of the chosen surfaces, which lie in the
coarse resolution region; the 'roughness' of the curves follows from
this. 

Figure~\ref{fig:am}, which refers to the piecewise-polytropic EoS, 
shows then the conservation of the angular
momentum, defined here as the sum of the angular momentum computed on
the numerical domain and of the angular momentum carried by 
gravitational waves outside the numerical domain. Such a quantity,
normalized to its initial value (the initial angular momentum)
should be constant and the figure shows the deviation of the results
from constancy. The colours and line types are the same as in the
previous figures. At the highest resolution, {\tt Whisky} conserves
this quantity very well, at better than 1\%, and also for {\tt
SACRA} deviations from constancy are of the same order, even if larger
oscillations are visible. The difference in the computation of the
angular momentum, analogous the one for the ADM mass, is also here at
the origin of the difference in the smoothness of the curves. Namely,
{\tt Whisky} performs a volume integral with the
formula~\cite{Shibata99c}
\begin{align}
\label{AngMomFormula}
J^i_{\rm vol} &=& \varepsilon^{ijk}\int_V \bigg(\frac{1}{8\pi}\tilde A_{jk}
+ x_j S_k + \frac{1}{12\pi}x_j K_{,k} + \nonumber \\
&&-\frac{1}{16\pi}x_j\tilde\gamma^{lm}{}_{,k}\tilde A_{lm}\bigg)
e^{6\phi} d^3x
\end{align} 
and excludes from the integral the points inside the AH. However, if
the angular momentum of the black hole is added to the one computed
above, the correct time evolution of the quantity in Fig.~\ref{fig:am}
is recovered, except for an interval just after the AH formation, when
the AH is small and covers only a few grid points, and so the
measurement of its angular momentum is inaccurate.
{\tt SACRA}, instead, uses also here a surface integral.

As previously noted, also from the time evolutions in
Fig.~\ref{fig:am} one sees that the conservation of the angular
momentum at these resolutions depends in a stronger way on resolution for the {\tt Whisky}
code with respect to {\tt SACRA}. In addition, one can see that, while also the {\tt SACRA} data
show convergence almost everywhere, in some time intervals the behaviour at different resolutions is not
convergent, for example at the spike around $2.5$ ms. The reason is not completely clear at the
moment, but we think that this is probably related to the low resolution of the coarsest grid, 
where the surface on which the angular momentum is computed is located [note that accurate extraction of 
angular momentum requires an accurate computation of parts of the extrinsic curvature that are
$O(r^{-3})$ and these are much smaller than the 
leading-order wave part of $O(r^{-1})$].
If the angular momentum
is computed on surfaces that lie on the finer levels, the differences in the wrong 
direction caused by resolution are much 
smaller (but the value of the angular momentum is less accurate).\\

We now proceed to analyze gravitational waves extracted from the
simulations. The data presented here are extracted from the numerical 
simulations at distances from the origin of the axes in
the interval $300\sim 600$ km.
For building templates to be used in the analysis of the
data taken by the gravitational-wave detectors, the accurate knowledge
of the frequency of the waves is of special importance. Thus we first
show in the left panel of Fig.~\ref{fig:freq}, which refers to the ideal-fluid EoS, the comparison of the
orbital frequency. The agreement of the results of the two codes is
excellent, if one ignores the initial spurious signal (related to the
spurious gravitational-wave content of the initial data, which is
rapidly propagated away). The orbital frequency $\Omega$ is computed in postprocessing from the
time derivatives of the real and imaginary part of $\psi_4$:
\begin{equation}
\Omega = -\frac{{\mathrm d}}{{\mathrm d} t} \bigg [{\rm atan}{\frac{\Im(\psi_4)}{\Re(\psi_4)}} \bigg ] =
- \frac{\frac{{\mathrm d}\Im(\psi_4)}{{\mathrm d}t}\Re(\psi_4) -\Im(\psi_4)\frac{{\mathrm
       d}\Re(\psi_4)}{{\mathrm d}t}} {[\Re(\psi_4)]^2+[\Im(\psi_4)]^2}.
\end{equation}
The biggest real ({\it i.e.} not related to the noise) difference between the two curves is during
the merger, at around $7.5$ ms and it is of about $10\%$, which is
consistent with the results of~\cite{Hannam:2009hh}.

In the right panel of Fig.~\ref{fig:freq} we plot the amplitude of 
waves as a function of the frequency. The inset shows that the error on
the amplitude is always at most $10\%$, which is of the same order of
magnitude of the one found in the comparison of numerical codes in binary--black-hole
simulations~\cite{Hannam:2009hh} and provides here an important
consistency check on the numerical accuracy and validity of the
waveforms of both {\tt Whisky} and {\tt SACRA}. The discussion of whether, as 
in~\cite{Hannam:2009hh}, also for binary-NS--merger waveforms this discrepancy 
is relevant or not for data analysis (namely whether current detectors can or
cannot distinguish between the waveforms of the two codes) is left to a future work, 
now in preparation~\cite{Baiotti:2010b}.

Finally, in order to give also a strong visual support to the goodness
of the consistence of gravitational waves computed from the two codes,
in Fig.~\ref{fig:h+}, which refers to the piecewise-polytropic EoS, 
we show the $(h_+)_{22}$ waveforms, together with the curve predicted by the 
Taylor-T4 post-Newtonian approximation~\cite{Baker-etal:2007b,Boyle:2007ft,Baiotti:2010b}. These are the
raw data, in the sense that no phase shift is performed to achieve the
best alignment of gravitational waves. The latter procedure is often
successfully performed in data-analysis related work and will be
included in our future work~\cite{Baiotti:2010b}. Nevertheless the similarity of the
numerical waveforms (both between themselves and with respect to the post-Newtonian prediction) 
in the inspiral part is astonishingly good at the highest
resolutions adopted here (see upper panel of Fig.~\ref{fig:h+}). On
the contrary, the lower resolutions (lower panel) are clearly not good
enough.

The situation is somewhat different after the merger. The ringdown part shows agreement between the
two codes, but the waves show some differences both in amplitude and frequency in the interval after
the merger and before the ring-down. This is due to the differences in the EoSs, as explained in
Sec.~\ref{sec:EoS}. Namely, in the present simulations {\tt SACRA} added a thermal part to the 
piecewise-polytropic EoS, while {\tt Whisky} did not. As shown by the figures, the difference in the
EoS are irrelevant to the inspiral phase, but not so after the merger, as expected.

\section{Conclusions}

In this work we have presented the first, detailed comparison of two
general-relativistic hydrodynamics codes, the {\tt Whisky} code and
the {\tt SACRA} code.

We have compared numerical-relativity waveforms and other quantities
for the last orbits, merger, and collapse of equal-mass irrotational
binary NS systems, as produced by the two independent computer
codes. We focused on two analytic EoSs, namely the simple ideal-fluid
EoS and a piecewise-polytropic EoS, for which we additionally
presented more resolutions. The purpose was to perform a stringent
consistency check of the results from these codes. We found that the
waveform frequency and amplitude computed with the two codes are in
agreement with a discrepancy of at most $10\%$ (this estimate refers to the merger time; the
discrepancy is much less during the inspiral), which is comparable to
the intrinsic error of each individual code at the adopted
resolutions. We stress the fact that this estimated error should be considered here an upper limit
and that the discrepancy between the waves computed in the two codes will be smaller when we will
consider an optimised overlap of the waveforms, in our future work~\cite{Baiotti:2010b}. 

The comparison of purely hydrodynamical quantities, like
the rest-mass density, shows better results, with a difference between the two codes of at most about
$1\%$. This number refers however only to global quantities (like maxima and norms), but not to
point-to-point comparisons, mainly because of the small phase difference in the evolution, which
makes pointwise comparisons meaningless. In fact, even after compensating for the
phase difference, errors larger than $1\%$ are seen at some points, noticeably those near the
surface of the stars. Such errors are related to different implementations of, {\it e.g.}, the
atmosphere treatment and do not influence the global dynamics in a noticeable way.

Finally, by comparing other time-dependent spacetime
and matter quantities, we showed that both codes conserve at high
accuracy rest mass, energy, and angular momentum, when taking into
account the emission of gravitational waves. The small differences that are
present have been related to details in the different implementations and grid setups.

In conclusion, encouraging results have been shown and more work is
now necessary to assess how the remaining differences in the results
may affect the construction of templates for gravitational-wave data
analysis. This will be the subject of a future work~\cite{Baiotti:2010b}, which may
include also more codes in the comparison.

\begin{acknowledgments}

  We are grateful to K. Taniguchi for providing the initial
  data we used in this work and to the Meudon group for producing and making public a code very
  useful to the community. LB thanks B. Giacomazzo, I. Hawke, J. Read, L. Rezzolla, 
  and E. Schnetter for useful comments and
  discussion, and acknowledges the continuous and great work done by
  the community developing the {\tt Cactus} and the {\tt Carpet}
  codes. The {\tt Whisky} simulations were performed on the Ranger
  cluster at the Texas Advanced Computing Center through TERAGRID
  allocation TG-MCA02N014 and at the CFCA of the National Astronomical
  Observatory of Japan. The work of LB was supported in part by the JSPS Postdoctoral Fellowship For Foreign
  Researchers, Grant-in-Aid for Scientific Research (19-07803), and by the Grant-in-Aid for Young Scientists
  (22740163); the work of MS was supported by the Grant-in-Aid
  for Scientific Research (21340051), and by the Grant-in-Aid for
  Scientific Research on Innovative Area (20105004) of the Japanese MEXT.
\end{acknowledgments}

\bibliographystyle{apsrev-nourl}
\bibliography{references}

\begin{thebibliography}{69}
\expandafter\ifx\csname natexlab\endcsname\relax\def\natexlab#1{#1}\fi
\expandafter\ifx\csname bibnamefont\endcsname\relax
  \def\bibnamefont#1{#1}\fi
\expandafter\ifx\csname bibfnamefont\endcsname\relax
  \def\bibfnamefont#1{#1}\fi
\expandafter\ifx\csname citenamefont\endcsname\relax
  \def\citenamefont#1{#1}\fi
\expandafter\ifx\csname url\endcsname\relax
  \def\url#1{\texttt{#1}}\fi
\expandafter\ifx\csname urlprefix\endcsname\relax\def\urlprefix{URL }\fi
\providecommand{\bibinfo}[2]{#2}
\providecommand{\eprint}[2][]{\url{#2}}

\bibitem[{\citenamefont{Baiotti
  et~al.}(2009{\natexlab{a}})\citenamefont{Baiotti, Giacomazzo, and
  Rezzolla}}]{Baiotti:2009gk}
\bibinfo{author}{\bibfnamefont{L.}~\bibnamefont{Baiotti}},
  \bibinfo{author}{\bibfnamefont{B.}~\bibnamefont{Giacomazzo}},
  \bibnamefont{and} \bibinfo{author}{\bibfnamefont{L.}~\bibnamefont{Rezzolla}},
  \bibinfo{journal}{Class. Quantum Grav.} \textbf{\bibinfo{volume}{26}},
  \bibinfo{pages}{114005} (\bibinfo{year}{2009}{\natexlab{a}}).

\bibitem[{\citenamefont{{Baiotti} et~al.}(2008)\citenamefont{{Baiotti},
  {Giacomazzo}, and {Rezzolla}}}]{Baiotti08}
\bibinfo{author}{\bibfnamefont{L.}~\bibnamefont{{Baiotti}}},
  \bibinfo{author}{\bibfnamefont{B.}~\bibnamefont{{Giacomazzo}}},
  \bibnamefont{and}
  \bibinfo{author}{\bibfnamefont{L.}~\bibnamefont{{Rezzolla}}},
  \bibinfo{journal}{Phys. Rev. D} \textbf{\bibinfo{volume}{78}},
  \bibinfo{pages}{084033} (\bibinfo{year}{2008}).

\bibitem[{\citenamefont{{Yamamoto} et~al.}(2008)\citenamefont{{Yamamoto},
  {Shibata}, and {Taniguchi}}}]{Yamamoto2008}
\bibinfo{author}{\bibfnamefont{T.}~\bibnamefont{{Yamamoto}}},
  \bibinfo{author}{\bibfnamefont{M.}~\bibnamefont{{Shibata}}},
  \bibnamefont{and}
  \bibinfo{author}{\bibfnamefont{K.}~\bibnamefont{{Taniguchi}}},
  \bibinfo{journal}{Phys. Rev. D} \textbf{\bibinfo{volume}{78}},
  \bibinfo{pages}{064054} (\bibinfo{year}{2008}).

\bibitem[{\citenamefont{Pretorius}(2005)}]{Pretorius:2005gq}
\bibinfo{author}{\bibfnamefont{F.}~\bibnamefont{Pretorius}},
  \bibinfo{journal}{Phys. Rev. Lett.} \textbf{\bibinfo{volume}{95}},
  \bibinfo{pages}{121101} (\bibinfo{year}{2005}).

\bibitem[{\citenamefont{Campanelli et~al.}(2006)\citenamefont{Campanelli,
  Lousto, Marronetti, and Zlochower}}]{Campanelli:2005dd}
\bibinfo{author}{\bibfnamefont{M.}~\bibnamefont{Campanelli}},
  \bibinfo{author}{\bibfnamefont{C.~O.} \bibnamefont{Lousto}},
  \bibinfo{author}{\bibfnamefont{P.}~\bibnamefont{Marronetti}},
  \bibnamefont{and}
  \bibinfo{author}{\bibfnamefont{Y.}~\bibnamefont{Zlochower}},
  \bibinfo{journal}{Phys. Rev. Lett.} \textbf{\bibinfo{volume}{96}},
  \bibinfo{pages}{111101} (\bibinfo{year}{2006}).

\bibitem[{\citenamefont{Baker et~al.}(2006{\natexlab{a}})\citenamefont{Baker,
  Centrella, Choi, Koppitz, and van Meter}}]{Baker:2005vv}
\bibinfo{author}{\bibfnamefont{J.~G.} \bibnamefont{Baker}},
  \bibinfo{author}{\bibfnamefont{J.}~\bibnamefont{Centrella}},
  \bibinfo{author}{\bibfnamefont{D.-I.} \bibnamefont{Choi}},
  \bibinfo{author}{\bibfnamefont{M.}~\bibnamefont{Koppitz}}, \bibnamefont{and}
  \bibinfo{author}{\bibfnamefont{J.}~\bibnamefont{van Meter}},
  \bibinfo{journal}{Phys. Rev. Lett.} \textbf{\bibinfo{volume}{96}},
  \bibinfo{pages}{111102} (\bibinfo{year}{2006}{\natexlab{a}}).

\bibitem[{\citenamefont{Hannam et~al.}(2009)}]{Hannam:2009hh}
\bibinfo{author}{\bibfnamefont{M.}~\bibnamefont{Hannam}} \bibnamefont{et~al.},
  \bibinfo{journal}{Phys. Rev. D} \textbf{\bibinfo{volume}{79}},
  \bibinfo{pages}{084025} (\bibinfo{year}{2009}).

\bibitem[{\citenamefont{Baker et~al.}(2007{\natexlab{a}})\citenamefont{Baker,
  Campanelli, Pretorius, and Zlochower}}]{Baker-etal:2007a}
\bibinfo{author}{\bibfnamefont{J.~G.} \bibnamefont{Baker}},
  \bibinfo{author}{\bibfnamefont{M.}~\bibnamefont{Campanelli}},
  \bibinfo{author}{\bibfnamefont{F.}~\bibnamefont{Pretorius}},
  \bibnamefont{and}
  \bibinfo{author}{\bibfnamefont{Y.}~\bibnamefont{Zlochower}},
  \bibinfo{journal}{Class. Quantum Grav.} \textbf{\bibinfo{volume}{24}},
  \bibinfo{pages}{S25} (\bibinfo{year}{2007}{\natexlab{a}}).

\bibitem[{\citenamefont{Br{\"u}gmann et~al.}(2008)\citenamefont{Br{\"u}gmann,
  Gonz{\'a}lez, Hannam, Husa, Sperhake, and Tichy}}]{Bruegmann:2006at}
\bibinfo{author}{\bibfnamefont{B.}~\bibnamefont{Br{\"u}gmann}},
  \bibinfo{author}{\bibfnamefont{J.~A.} \bibnamefont{Gonz{\'a}lez}},
  \bibinfo{author}{\bibfnamefont{M.}~\bibnamefont{Hannam}},
  \bibinfo{author}{\bibfnamefont{S.}~\bibnamefont{Husa}},
  \bibinfo{author}{\bibfnamefont{U.}~\bibnamefont{Sperhake}}, \bibnamefont{and}
  \bibinfo{author}{\bibfnamefont{W.}~\bibnamefont{Tichy}},
  \bibinfo{journal}{Phys.Rev.} \textbf{\bibinfo{volume}{D77}},
  \bibinfo{pages}{024027} (\bibinfo{year}{2008}),
  \bibinfo{note}{gr-qc/0610128}.

\bibitem[{LIGO()}]{LIGO_web}
LIGO, \bibinfo{note}{lIGO -- http://www.ligo.caltech.edu/}.

\bibitem[{VIRGO()}]{VIRGO_web}
VIRGO, \bibinfo{note}{vIRGO -- http://www.virgo.infn.it/}.

\bibitem[{GEO()}]{GEO_web}
GEO, \bibinfo{note}{gEO600 -- http://www.geo600.uni-hannover.de/}.

\bibitem[{\citenamefont{Punturo et~al.}(2010)}]{Punturo:2010}
\bibinfo{author}{\bibfnamefont{M.}~\bibnamefont{Punturo}} \bibnamefont{et~al.},
  \bibinfo{journal}{Class. Quantum Grav.} \textbf{\bibinfo{volume}{27}},
  \bibinfo{pages}{084007} (\bibinfo{year}{2010}).

\bibitem[{\citenamefont{{Einstein Telescope}}()}]{ET_noise}
\bibinfo{author}{\bibnamefont{{Einstein Telescope}}},
  \urlprefix\url{http://www.et-gw.eu}.

\bibitem[{\citenamefont{Baiotti et~al.}(2003)\citenamefont{Baiotti, Hawke,
  Montero, and Rezzolla}}]{Baiotti03a}
\bibinfo{author}{\bibfnamefont{L.}~\bibnamefont{Baiotti}},
  \bibinfo{author}{\bibfnamefont{I.}~\bibnamefont{Hawke}},
  \bibinfo{author}{\bibfnamefont{P.}~\bibnamefont{Montero}}, \bibnamefont{and}
  \bibinfo{author}{\bibfnamefont{L.}~\bibnamefont{Rezzolla}}, in
  \emph{\bibinfo{booktitle}{Computational Astrophysics in Italy: Methods and
  Tools}}, edited by
  \bibinfo{editor}{\bibfnamefont{R.}~\bibnamefont{Capuzzo-Dolcetta}}
  (\bibinfo{publisher}{MSAIt}, \bibinfo{address}{Trieste},
  \bibinfo{year}{2003}), vol.~\bibinfo{volume}{1}, p. \bibinfo{pages}{210}.

\bibitem[{\citenamefont{Baiotti et~al.}(2005)\citenamefont{Baiotti, Hawke,
  Montero, L{\"o}ffler, Rezzolla, Stergioulas, Font, and Seidel}}]{Baiotti04}
\bibinfo{author}{\bibfnamefont{L.}~\bibnamefont{Baiotti}},
  \bibinfo{author}{\bibfnamefont{I.}~\bibnamefont{Hawke}},
  \bibinfo{author}{\bibfnamefont{P.~J.} \bibnamefont{Montero}},
  \bibinfo{author}{\bibfnamefont{F.}~\bibnamefont{L{\"o}ffler}},
  \bibinfo{author}{\bibfnamefont{L.}~\bibnamefont{Rezzolla}},
  \bibinfo{author}{\bibfnamefont{N.}~\bibnamefont{Stergioulas}},
  \bibinfo{author}{\bibfnamefont{J.~A.} \bibnamefont{Font}}, \bibnamefont{and}
  \bibinfo{author}{\bibfnamefont{E.}~\bibnamefont{Seidel}},
  \bibinfo{journal}{Phys. Rev. D} \textbf{\bibinfo{volume}{71}},
  \bibinfo{pages}{024035} (\bibinfo{year}{2005}).

\bibitem[{\citenamefont{Giacomazzo and Rezzolla}(2007)}]{Giacomazzo:2007ti}
\bibinfo{author}{\bibfnamefont{B.}~\bibnamefont{Giacomazzo}} \bibnamefont{and}
  \bibinfo{author}{\bibfnamefont{L.}~\bibnamefont{Rezzolla}},
  \bibinfo{journal}{Class. Quantum Grav.} \textbf{\bibinfo{volume}{24}},
  \bibinfo{pages}{S235} (\bibinfo{year}{2007}).

\bibitem[{\citenamefont{Baiotti et~al.}(2010)\citenamefont{Baiotti, Creighton,
  Friedman, Giacomazzo, Markakis, Read, Rezzolla, Shibata, Taniguchi
  et~al.}}]{Baiotti:2010b}
\bibinfo{author}{\bibfnamefont{L.}~\bibnamefont{Baiotti}},
  \bibinfo{author}{\bibfnamefont{J.~D.~E.} \bibnamefont{Creighton}},
  \bibinfo{author}{\bibfnamefont{J.~L.} \bibnamefont{Friedman}},
  \bibinfo{author}{\bibfnamefont{B.}~\bibnamefont{Giacomazzo}},
  \bibinfo{author}{\bibfnamefont{C.}~\bibnamefont{Markakis}},
  \bibinfo{author}{\bibfnamefont{J.~S.} \bibnamefont{Read}},
  \bibinfo{author}{\bibfnamefont{L.}~\bibnamefont{Rezzolla}},
  \bibinfo{author}{\bibfnamefont{M.}~\bibnamefont{Shibata}},
  \bibinfo{author}{\bibfnamefont{K.}~\bibnamefont{Taniguchi}},
  \bibnamefont{et~al.}, \bibinfo{journal}{in preparation}
  (\bibinfo{year}{2010}).

\bibitem[{\citenamefont{Nakar}(2007)}]{Nakar:2007yr}
\bibinfo{author}{\bibfnamefont{E.}~\bibnamefont{Nakar}},
  \bibinfo{journal}{Phys. Rep.} \textbf{\bibinfo{volume}{442}},
  \bibinfo{pages}{166} (\bibinfo{year}{2007}).

\bibitem[{\citenamefont{Pollney et~al.}(2007)\citenamefont{Pollney, Reisswig,
  Rezzolla, Szil{\'a}gyi, Ansorg, Deris, Diener, Dorband, Koppitz, Nagar
  et~al.}}]{Pollney:2007ss}
\bibinfo{author}{\bibfnamefont{D.}~\bibnamefont{Pollney}},
  \bibinfo{author}{\bibfnamefont{C.}~\bibnamefont{Reisswig}},
  \bibinfo{author}{\bibfnamefont{L.}~\bibnamefont{Rezzolla}},
  \bibinfo{author}{\bibfnamefont{B.}~\bibnamefont{Szil{\'a}gyi}},
  \bibinfo{author}{\bibfnamefont{M.}~\bibnamefont{Ansorg}},
  \bibinfo{author}{\bibfnamefont{B.}~\bibnamefont{Deris}},
  \bibinfo{author}{\bibfnamefont{P.}~\bibnamefont{Diener}},
  \bibinfo{author}{\bibfnamefont{E.~N.} \bibnamefont{Dorband}},
  \bibinfo{author}{\bibfnamefont{M.}~\bibnamefont{Koppitz}},
  \bibinfo{author}{\bibfnamefont{A.}~\bibnamefont{Nagar}},
  \bibnamefont{et~al.}, \bibinfo{journal}{Phys. Rev. D}
  \textbf{\bibinfo{volume}{76}}, \bibinfo{pages}{124002}
  (\bibinfo{year}{2007}).

\bibitem[{\citenamefont{Nakamura et~al.}(1987)\citenamefont{Nakamura, Oohara,
  and Kojima}}]{Nakamura87}
\bibinfo{author}{\bibfnamefont{T.}~\bibnamefont{Nakamura}},
  \bibinfo{author}{\bibfnamefont{K.}~\bibnamefont{Oohara}}, \bibnamefont{and}
  \bibinfo{author}{\bibfnamefont{Y.}~\bibnamefont{Kojima}},
  \bibinfo{journal}{Prog. Theor. Phys. Suppl.} \textbf{\bibinfo{volume}{90}},
  \bibinfo{pages}{1} (\bibinfo{year}{1987}).

\bibitem[{\citenamefont{Shibata and Nakamura}(1995)}]{Shibata95}
\bibinfo{author}{\bibfnamefont{M.}~\bibnamefont{Shibata}} \bibnamefont{and}
  \bibinfo{author}{\bibfnamefont{T.}~\bibnamefont{Nakamura}},
  \bibinfo{journal}{Phys. Rev. D} \textbf{\bibinfo{volume}{52}},
  \bibinfo{pages}{5428} (\bibinfo{year}{1995}).

\bibitem[{\citenamefont{Baumgarte and Shapiro}(1998)}]{Baumgarte99}
\bibinfo{author}{\bibfnamefont{T.~W.} \bibnamefont{Baumgarte}}
  \bibnamefont{and} \bibinfo{author}{\bibfnamefont{S.~L.}
  \bibnamefont{Shapiro}}, \bibinfo{journal}{Phys. Rev. D}
  \textbf{\bibinfo{volume}{59}}, \bibinfo{pages}{024007}
  (\bibinfo{year}{1998}).

\bibitem[{\citenamefont{Alcubierre et~al.}(2000)\citenamefont{Alcubierre,
  Br{\"u}gmann, Dramlitsch, Font, Papadopoulos, Seidel, Stergioulas, and
  Takahashi}}]{Alcubierre99d}
\bibinfo{author}{\bibfnamefont{M.}~\bibnamefont{Alcubierre}},
  \bibinfo{author}{\bibfnamefont{B.}~\bibnamefont{Br{\"u}gmann}},
  \bibinfo{author}{\bibfnamefont{T.}~\bibnamefont{Dramlitsch}},
  \bibinfo{author}{\bibfnamefont{J.~A.} \bibnamefont{Font}},
  \bibinfo{author}{\bibfnamefont{P.}~\bibnamefont{Papadopoulos}},
  \bibinfo{author}{\bibfnamefont{E.}~\bibnamefont{Seidel}},
  \bibinfo{author}{\bibfnamefont{N.}~\bibnamefont{Stergioulas}},
  \bibnamefont{and}
  \bibinfo{author}{\bibfnamefont{R.}~\bibnamefont{Takahashi}},
  \bibinfo{journal}{Phys. Rev. D} \textbf{\bibinfo{volume}{62}},
  \bibinfo{pages}{044034} (\bibinfo{year}{2000}).

\bibitem[{\citenamefont{Goodale et~al.}(2003)\citenamefont{Goodale, Allen,
  Lanfermann, Mass{\'o}, Radke, Seidel, and Shalf}}]{Goodale02a}
\bibinfo{author}{\bibfnamefont{T.}~\bibnamefont{Goodale}},
  \bibinfo{author}{\bibfnamefont{G.}~\bibnamefont{Allen}},
  \bibinfo{author}{\bibfnamefont{G.}~\bibnamefont{Lanfermann}},
  \bibinfo{author}{\bibfnamefont{J.}~\bibnamefont{Mass{\'o}}},
  \bibinfo{author}{\bibfnamefont{T.}~\bibnamefont{Radke}},
  \bibinfo{author}{\bibfnamefont{E.}~\bibnamefont{Seidel}}, \bibnamefont{and}
  \bibinfo{author}{\bibfnamefont{J.}~\bibnamefont{Shalf}}, in
  \emph{\bibinfo{booktitle}{Vector and Parallel Processing -- VECPAR'2002, 5th
  International Conference, Lecture Notes in Computer Science}}
  (\bibinfo{publisher}{Springer}, \bibinfo{address}{Berlin},
  \bibinfo{year}{2003}).

\bibitem[{\citenamefont{York}(1979)}]{York79}
\bibinfo{author}{\bibfnamefont{J.~W.} \bibnamefont{York}}, in
  \emph{\bibinfo{booktitle}{Sources of gravitational radiation}}, edited by
  \bibinfo{editor}{\bibfnamefont{L.~L.} \bibnamefont{Smarr}}
  (\bibinfo{publisher}{Cambridge University Press},
  \bibinfo{address}{Cambridge, UK}, \bibinfo{year}{1979}), pp.
  \bibinfo{pages}{83--126}, ISBN \bibinfo{isbn}{0-521-22778-X}.

\bibitem[{\citenamefont{Alcubierre et~al.}(2003)\citenamefont{Alcubierre,
  Br{\"u}gmann, Diener, Koppitz, Pollney, Seidel, and
  Takahashi}}]{Alcubierre02a}
\bibinfo{author}{\bibfnamefont{M.}~\bibnamefont{Alcubierre}},
  \bibinfo{author}{\bibfnamefont{B.}~\bibnamefont{Br{\"u}gmann}},
  \bibinfo{author}{\bibfnamefont{P.}~\bibnamefont{Diener}},
  \bibinfo{author}{\bibfnamefont{M.}~\bibnamefont{Koppitz}},
  \bibinfo{author}{\bibfnamefont{D.}~\bibnamefont{Pollney}},
  \bibinfo{author}{\bibfnamefont{E.}~\bibnamefont{Seidel}}, \bibnamefont{and}
  \bibinfo{author}{\bibfnamefont{R.}~\bibnamefont{Takahashi}},
  \bibinfo{journal}{Phys. Rev. D} \textbf{\bibinfo{volume}{67}},
  \bibinfo{pages}{084023} (\bibinfo{year}{2003}).

\bibitem[{\citenamefont{Misner et~al.}(1973)\citenamefont{Misner, Thorne, and
  Wheeler}}]{misner73}
\bibinfo{author}{\bibfnamefont{C.~W.} \bibnamefont{Misner}},
  \bibinfo{author}{\bibfnamefont{K.~S.} \bibnamefont{Thorne}},
  \bibnamefont{and} \bibinfo{author}{\bibfnamefont{J.~A.}
  \bibnamefont{Wheeler}}, \emph{\bibinfo{title}{Gravitation}}
  (\bibinfo{publisher}{W. H. Freeman}, \bibinfo{address}{San Francisco},
  \bibinfo{year}{1973}).

\bibitem[{\citenamefont{Bona et~al.}(1995)\citenamefont{Bona, Mass{\'o},
  Seidel, and Stela}}]{Bona94b}
\bibinfo{author}{\bibfnamefont{C.}~\bibnamefont{Bona}},
  \bibinfo{author}{\bibfnamefont{J.}~\bibnamefont{Mass{\'o}}},
  \bibinfo{author}{\bibfnamefont{E.}~\bibnamefont{Seidel}}, \bibnamefont{and}
  \bibinfo{author}{\bibfnamefont{J.}~\bibnamefont{Stela}},
  \bibinfo{journal}{Phys. Rev. Lett.} \textbf{\bibinfo{volume}{75}},
  \bibinfo{pages}{600} (\bibinfo{year}{1995}).

\bibitem[{\citenamefont{Baker et~al.}(2006{\natexlab{b}})\citenamefont{Baker,
  Centrella, Choi, Koppitz, and van Meter}}]{Baker05a}
\bibinfo{author}{\bibfnamefont{J.~G.} \bibnamefont{Baker}},
  \bibinfo{author}{\bibfnamefont{J.}~\bibnamefont{Centrella}},
  \bibinfo{author}{\bibfnamefont{D.-I.} \bibnamefont{Choi}},
  \bibinfo{author}{\bibfnamefont{M.}~\bibnamefont{Koppitz}}, \bibnamefont{and}
  \bibinfo{author}{\bibfnamefont{J.}~\bibnamefont{van Meter}},
  \bibinfo{journal}{Phys. Rev. Lett.} \textbf{\bibinfo{volume}{96}},
  \bibinfo{pages}{111102} (\bibinfo{year}{2006}{\natexlab{b}}).

\bibitem[{\citenamefont{van Meter et~al.}(2006)\citenamefont{van Meter, Baker,
  Koppitz, and Choi}}]{Baker:2006mp}
\bibinfo{author}{\bibfnamefont{J.}~\bibnamefont{van Meter}},
  \bibinfo{author}{\bibfnamefont{J.~G.} \bibnamefont{Baker}},
  \bibinfo{author}{\bibfnamefont{M.}~\bibnamefont{Koppitz}}, \bibnamefont{and}
  \bibinfo{author}{\bibfnamefont{D.-I.} \bibnamefont{Choi}},
  \bibinfo{journal}{Phys. Rev. D} \textbf{\bibinfo{volume}{73}},
  \bibinfo{pages}{124011} (\bibinfo{year}{2006}).

\bibitem[{\citenamefont{Koppitz et~al.}(2007)\citenamefont{Koppitz, Pollney,
  Reisswig, Rezzolla, Thornburg, Diener, and Schnetter}}]{Koppitz-etal-2007aa}
\bibinfo{author}{\bibfnamefont{M.}~\bibnamefont{Koppitz}},
  \bibinfo{author}{\bibfnamefont{D.}~\bibnamefont{Pollney}},
  \bibinfo{author}{\bibfnamefont{C.}~\bibnamefont{Reisswig}},
  \bibinfo{author}{\bibfnamefont{L.}~\bibnamefont{Rezzolla}},
  \bibinfo{author}{\bibfnamefont{J.}~\bibnamefont{Thornburg}},
  \bibinfo{author}{\bibfnamefont{P.}~\bibnamefont{Diener}}, \bibnamefont{and}
  \bibinfo{author}{\bibfnamefont{E.}~\bibnamefont{Schnetter}},
  \bibinfo{journal}{Phys. Rev. Lett.} \textbf{\bibinfo{volume}{99}},
  \bibinfo{pages}{041102} (\bibinfo{year}{2007}).

\bibitem[{\citenamefont{Thornburg}(2004)}]{Thornburg2003:AH-finding_nourl}
\bibinfo{author}{\bibfnamefont{J.}~\bibnamefont{Thornburg}},
  \bibinfo{journal}{Class. Quantum Grav.} \textbf{\bibinfo{volume}{21}},
  \bibinfo{pages}{743} (\bibinfo{year}{2004}).

\bibitem[{\citenamefont{Thornburg}(1996)}]{Thornburg95}
\bibinfo{author}{\bibfnamefont{J.}~\bibnamefont{Thornburg}},
  \bibinfo{journal}{Phys. Rev. D} \textbf{\bibinfo{volume}{54}},
  \bibinfo{pages}{4899} (\bibinfo{year}{1996}).

\bibitem[{\citenamefont{Ashtekar
  et~al.}(2000{\natexlab{a}})\citenamefont{Ashtekar, Beetle, and
  Fairhurst}}]{Ashtekar99a}
\bibinfo{author}{\bibfnamefont{A.}~\bibnamefont{Ashtekar}},
  \bibinfo{author}{\bibfnamefont{C.}~\bibnamefont{Beetle}}, \bibnamefont{and}
  \bibinfo{author}{\bibfnamefont{S.}~\bibnamefont{Fairhurst}},
  \bibinfo{journal}{Class. Quantum Grav.} \textbf{\bibinfo{volume}{17}},
  \bibinfo{pages}{253} (\bibinfo{year}{2000}{\natexlab{a}}).

\bibitem[{\citenamefont{Ashtekar
  et~al.}(2000{\natexlab{b}})\citenamefont{Ashtekar, Beetle, Dreyer, Fairhurst,
  Krishnan, Lewandowski, and Wisniewski}}]{Ashtekar00a}
\bibinfo{author}{\bibfnamefont{A.}~\bibnamefont{Ashtekar}},
  \bibinfo{author}{\bibfnamefont{C.}~\bibnamefont{Beetle}},
  \bibinfo{author}{\bibfnamefont{O.}~\bibnamefont{Dreyer}},
  \bibinfo{author}{\bibfnamefont{S.}~\bibnamefont{Fairhurst}},
  \bibinfo{author}{\bibfnamefont{B.}~\bibnamefont{Krishnan}},
  \bibinfo{author}{\bibfnamefont{J.}~\bibnamefont{Lewandowski}},
  \bibnamefont{and}
  \bibinfo{author}{\bibfnamefont{J.}~\bibnamefont{Wisniewski}},
  \bibinfo{journal}{Phys. Rev. Lett.} \textbf{\bibinfo{volume}{85}},
  \bibinfo{pages}{3564} (\bibinfo{year}{2000}{\natexlab{b}}).

\bibitem[{\citenamefont{Ashtekar et~al.}(2001)\citenamefont{Ashtekar, Beetle,
  and Lewandowski}}]{Ashtekar01a}
\bibinfo{author}{\bibfnamefont{A.}~\bibnamefont{Ashtekar}},
  \bibinfo{author}{\bibfnamefont{C.}~\bibnamefont{Beetle}}, \bibnamefont{and}
  \bibinfo{author}{\bibfnamefont{J.}~\bibnamefont{Lewandowski}},
  \bibinfo{journal}{Phys. Rev. D} \textbf{\bibinfo{volume}{64}},
  \bibinfo{pages}{044016} (\bibinfo{year}{2001}).

\bibitem[{\citenamefont{Ashtekar and
  Krishnan}(2002)}]{Ashtekar-etal-2002-dynamical-horizons}
\bibinfo{author}{\bibfnamefont{A.}~\bibnamefont{Ashtekar}} \bibnamefont{and}
  \bibinfo{author}{\bibfnamefont{B.}~\bibnamefont{Krishnan}},
  \bibinfo{journal}{Phys. Rev. Lett.} \textbf{\bibinfo{volume}{89}},
  \bibinfo{pages}{261101} (\bibinfo{year}{2002}).

\bibitem[{\citenamefont{Dreyer et~al.}(2003)\citenamefont{Dreyer, Krishnan,
  Shoemaker, and Schnetter}}]{Dreyer-etal-2002-isolated-horizons}
\bibinfo{author}{\bibfnamefont{O.}~\bibnamefont{Dreyer}},
  \bibinfo{author}{\bibfnamefont{B.}~\bibnamefont{Krishnan}},
  \bibinfo{author}{\bibfnamefont{D.}~\bibnamefont{Shoemaker}},
  \bibnamefont{and}
  \bibinfo{author}{\bibfnamefont{E.}~\bibnamefont{Schnetter}},
  \bibinfo{journal}{Phys. Rev. D} \textbf{\bibinfo{volume}{67}},
  \bibinfo{pages}{024018} (\bibinfo{year}{2003}).

\bibitem[{\citenamefont{Gunnarsen et~al.}(1995)\citenamefont{Gunnarsen,
  Shinkai, and Maeda}}]{Shinkai94}
\bibinfo{author}{\bibfnamefont{L.}~\bibnamefont{Gunnarsen}},
  \bibinfo{author}{\bibfnamefont{H.}~\bibnamefont{Shinkai}}, \bibnamefont{and}
  \bibinfo{author}{\bibfnamefont{K.}~\bibnamefont{Maeda}},
  \bibinfo{journal}{Class. Quantum Grav.} \textbf{\bibinfo{volume}{12}},
  \bibinfo{pages}{133} (\bibinfo{year}{1995}).

\bibitem[{\citenamefont{Teukolsky}(1973)}]{Teukolsky73}
\bibinfo{author}{\bibfnamefont{S.~A.} \bibnamefont{Teukolsky}},
  \bibinfo{journal}{Astrophys. J.} \textbf{\bibinfo{volume}{185}},
  \bibinfo{pages}{635} (\bibinfo{year}{1973}).

\bibitem[{\citenamefont{Baiotti
  et~al.}(2009{\natexlab{b}})\citenamefont{Baiotti, Bernuzzi, Corvino,
  De~Pietri, and Nagar}}]{Baiotti:2008nf}
\bibinfo{author}{\bibfnamefont{L.}~\bibnamefont{Baiotti}},
  \bibinfo{author}{\bibfnamefont{S.}~\bibnamefont{Bernuzzi}},
  \bibinfo{author}{\bibfnamefont{G.}~\bibnamefont{Corvino}},
  \bibinfo{author}{\bibfnamefont{R.}~\bibnamefont{De~Pietri}},
  \bibnamefont{and} \bibinfo{author}{\bibfnamefont{A.}~\bibnamefont{Nagar}},
  \bibinfo{journal}{Phys. Rev. D} \textbf{\bibinfo{volume}{79}},
  \bibinfo{pages}{024002} (\bibinfo{year}{2009}{\natexlab{b}}).

\bibitem[{\citenamefont{Moncrief}(1974)}]{Moncrief74}
\bibinfo{author}{\bibfnamefont{V.}~\bibnamefont{Moncrief}},
  \bibinfo{journal}{Annals of Physics} \textbf{\bibinfo{volume}{88}},
  \bibinfo{pages}{323} (\bibinfo{year}{1974}).

\bibitem[{\citenamefont{Mart{\'\i} et~al.}(1991)\citenamefont{Mart{\'\i},
  Ib{\'a}{\~n}ez, and Miralles}}]{Marti91}
\bibinfo{author}{\bibfnamefont{J.~M.} \bibnamefont{Mart{\'\i}}},
  \bibinfo{author}{\bibfnamefont{J.~M.} \bibnamefont{Ib{\'a}{\~n}ez}},
  \bibnamefont{and} \bibinfo{author}{\bibfnamefont{J.~A.}
  \bibnamefont{Miralles}}, \bibinfo{journal}{Phys. Rev. D}
  \textbf{\bibinfo{volume}{43}}, \bibinfo{pages}{3794} (\bibinfo{year}{1991}).

\bibitem[{\citenamefont{Banyuls et~al.}(1997)\citenamefont{Banyuls, Font,
  Ib{\'a}{\~n}ez, Mart{\'\i}, and Miralles}}]{Banyuls97}
\bibinfo{author}{\bibfnamefont{F.}~\bibnamefont{Banyuls}},
  \bibinfo{author}{\bibfnamefont{J.~A.} \bibnamefont{Font}},
  \bibinfo{author}{\bibfnamefont{J.~M.} \bibnamefont{Ib{\'a}{\~n}ez}},
  \bibinfo{author}{\bibfnamefont{J.~M.} \bibnamefont{Mart{\'\i}}},
  \bibnamefont{and} \bibinfo{author}{\bibfnamefont{J.~A.}
  \bibnamefont{Miralles}}, \bibinfo{journal}{Astrophys. J.}
  \textbf{\bibinfo{volume}{476}}, \bibinfo{pages}{221} (\bibinfo{year}{1997}).

\bibitem[{\citenamefont{Ib{\'a}{\~n}ez
  et~al.}(2001)\citenamefont{Ib{\'a}{\~n}ez, Aloy, Font, Mart{\'\i}, Miralles,
  and Pons}}]{Ibanez01}
\bibinfo{author}{\bibfnamefont{J.}~\bibnamefont{Ib{\'a}{\~n}ez}},
  \bibinfo{author}{\bibfnamefont{M.}~\bibnamefont{Aloy}},
  \bibinfo{author}{\bibfnamefont{J.}~\bibnamefont{Font}},
  \bibinfo{author}{\bibfnamefont{J.}~\bibnamefont{Mart{\'\i}}},
  \bibinfo{author}{\bibfnamefont{J.}~\bibnamefont{Miralles}}, \bibnamefont{and}
  \bibinfo{author}{\bibfnamefont{J.}~\bibnamefont{Pons}}, in
  \emph{\bibinfo{booktitle}{Godunov methods: theory and applications}}, edited
  by \bibinfo{editor}{\bibfnamefont{E.}~\bibnamefont{Toro}}
  (\bibinfo{publisher}{Kluwer Academic/Plenum Publishers},
  \bibinfo{address}{New York}, \bibinfo{year}{2001}).

\bibitem[{\citenamefont{Font}(2003)}]{Font03}
\bibinfo{author}{\bibfnamefont{J.~A.} \bibnamefont{Font}},
  \bibinfo{journal}{Living Rev. Relativ.} \textbf{\bibinfo{volume}{6}},
  \bibinfo{pages}{4} (\bibinfo{year}{2003}).

\bibitem[{\citenamefont{Toro}(1999)}]{Toro99}
\bibinfo{author}{\bibfnamefont{E.~F.} \bibnamefont{Toro}},
  \emph{\bibinfo{title}{Riemann Solvers and Numerical Methods for Fluid
  Dynamics}} (\bibinfo{publisher}{Springer-Verlag}, \bibinfo{year}{1999}).

\bibitem[{\citenamefont{Teukolsky}(2000)}]{Teukolsky00}
\bibinfo{author}{\bibfnamefont{S.~A.} \bibnamefont{Teukolsky}},
  \bibinfo{journal}{Phys. Rev. D} \textbf{\bibinfo{volume}{61}},
  \bibinfo{pages}{087501} (\bibinfo{year}{2000}).

\bibitem[{\citenamefont{Leiler and Rezzolla}(2006)}]{Leiler_Rezzolla06}
\bibinfo{author}{\bibfnamefont{G.}~\bibnamefont{Leiler}} \bibnamefont{and}
  \bibinfo{author}{\bibfnamefont{L.}~\bibnamefont{Rezzolla}},
  \bibinfo{journal}{Phys. Rev. D} \textbf{\bibinfo{volume}{73}},
  \bibinfo{pages}{044001} (\bibinfo{year}{2006}).

\bibitem[{\citenamefont{Colella and Woodward}(1984)}]{Colella84}
\bibinfo{author}{\bibfnamefont{P.}~\bibnamefont{Colella}} \bibnamefont{and}
  \bibinfo{author}{\bibfnamefont{P.~R.} \bibnamefont{Woodward}},
  \bibinfo{journal}{J. Comput. Phys.} \textbf{\bibinfo{volume}{54}},
  \bibinfo{pages}{174} (\bibinfo{year}{1984}).

\bibitem[{\citenamefont{Kurganov and Tadmor.}(2000)}]{Tadmor00}
\bibinfo{author}{\bibfnamefont{A.}~\bibnamefont{Kurganov}} \bibnamefont{and}
  \bibinfo{author}{\bibfnamefont{E.}~\bibnamefont{Tadmor.}},
  \bibinfo{journal}{J. Comput. Phys.} \textbf{\bibinfo{volume}{160}},
  \bibinfo{pages}{241} (\bibinfo{year}{2000}).

\bibitem[{\citenamefont{Harten et~al.}(1983)\citenamefont{Harten, Lax, and van
  Leer}}]{Harten83}
\bibinfo{author}{\bibfnamefont{A.}~\bibnamefont{Harten}},
  \bibinfo{author}{\bibfnamefont{P.~D.} \bibnamefont{Lax}}, \bibnamefont{and}
  \bibinfo{author}{\bibfnamefont{B.}~\bibnamefont{van Leer}},
  \bibinfo{journal}{SIAM Rev.} \textbf{\bibinfo{volume}{25}},
  \bibinfo{pages}{35} (\bibinfo{year}{1983}).

\bibitem[{\citenamefont{Aloy et~al.}(1999)\citenamefont{Aloy, Ib{\'a}{\~n}ez,
  Mart{\'\i}, and M{\"u}ller}}]{Aloy99b}
\bibinfo{author}{\bibfnamefont{M.~A.} \bibnamefont{Aloy}},
  \bibinfo{author}{\bibfnamefont{J.~M.} \bibnamefont{Ib{\'a}{\~n}ez}},
  \bibinfo{author}{\bibfnamefont{J.~M.} \bibnamefont{Mart{\'\i}}},
  \bibnamefont{and}
  \bibinfo{author}{\bibfnamefont{E.}~\bibnamefont{M{\"u}ller}},
  \bibinfo{journal}{Astrophys. J. Supp.} \textbf{\bibinfo{volume}{122}},
  \bibinfo{pages}{151} (\bibinfo{year}{1999}).

\bibitem[{\citenamefont{{Giacomazzo} et~al.}(2009)\citenamefont{{Giacomazzo},
  {Rezzolla}, and {Baiotti}}}]{Giacomazzo:2009mp}
\bibinfo{author}{\bibfnamefont{B.}~\bibnamefont{{Giacomazzo}}},
  \bibinfo{author}{\bibfnamefont{L.}~\bibnamefont{{Rezzolla}}},
  \bibnamefont{and}
  \bibinfo{author}{\bibfnamefont{L.}~\bibnamefont{{Baiotti}}},
  \bibinfo{journal}{MNRAS} \textbf{\bibinfo{volume}{399}},
  \bibinfo{pages}{L164} (\bibinfo{year}{2009}).

\bibitem[{\citenamefont{{Read} et~al.}(2009{\natexlab{a}})\citenamefont{{Read},
  {Lackey}, {Owen}, and {Friedman}}}]{Read:2009a}
\bibinfo{author}{\bibfnamefont{J.~S.} \bibnamefont{{Read}}},
  \bibinfo{author}{\bibfnamefont{B.~D.} \bibnamefont{{Lackey}}},
  \bibinfo{author}{\bibfnamefont{B.~J.} \bibnamefont{{Owen}}},
  \bibnamefont{and} \bibinfo{author}{\bibfnamefont{J.~L.}
  \bibnamefont{{Friedman}}}, \bibinfo{journal}{Phys. Rev. D}
  \textbf{\bibinfo{volume}{79}}, \bibinfo{pages}{124032}
  (\bibinfo{year}{2009}{\natexlab{a}}).

\bibitem[{\citenamefont{{Read} et~al.}(2009{\natexlab{b}})\citenamefont{{Read},
  {Markakis}, {Shibata}, {Uryu}, {Creighton}, and {Friedman}}}]{Read:2009b}
\bibinfo{author}{\bibfnamefont{J.~S.} \bibnamefont{{Read}}},
  \bibinfo{author}{\bibfnamefont{C.}~\bibnamefont{{Markakis}}},
  \bibinfo{author}{\bibfnamefont{M.}~\bibnamefont{{Shibata}}},
  \bibinfo{author}{\bibfnamefont{K.}~\bibnamefont{{Uryu}}},
  \bibinfo{author}{\bibfnamefont{J.~D.~E.} \bibnamefont{{Creighton}}},
  \bibnamefont{and} \bibinfo{author}{\bibfnamefont{J.~L.}
  \bibnamefont{{Friedman}}}, \bibinfo{journal}{Phys. Rev. D}
  \textbf{\bibinfo{volume}{79}}, \bibinfo{pages}{124033}
  (\bibinfo{year}{2009}{\natexlab{b}}).

\bibitem[{\citenamefont{Berger and Oliger}(1984)}]{Berger84}
\bibinfo{author}{\bibfnamefont{M.~J.} \bibnamefont{Berger}} \bibnamefont{and}
  \bibinfo{author}{\bibfnamefont{J.}~\bibnamefont{Oliger}},
  \bibinfo{journal}{J. Comput. Phys.} \textbf{\bibinfo{volume}{53}},
  \bibinfo{pages}{484} (\bibinfo{year}{1984}).

\bibitem[{\citenamefont{Schnetter et~al.}(2004)\citenamefont{Schnetter, Hawley,
  and Hawke}}]{Schnetter-etal-03b}
\bibinfo{author}{\bibfnamefont{E.}~\bibnamefont{Schnetter}},
  \bibinfo{author}{\bibfnamefont{S.~H.} \bibnamefont{Hawley}},
  \bibnamefont{and} \bibinfo{author}{\bibfnamefont{I.}~\bibnamefont{Hawke}},
  \bibinfo{journal}{Class. Quantum Grav.} \textbf{\bibinfo{volume}{21}},
  \bibinfo{pages}{1465} (\bibinfo{year}{2004}).

\bibitem[{\citenamefont{Harten et~al.}(1987)\citenamefont{Harten, Engquist,
  Osher, and Chakrabarty}}]{Harten87}
\bibinfo{author}{\bibfnamefont{A.}~\bibnamefont{Harten}},
  \bibinfo{author}{\bibfnamefont{B.}~\bibnamefont{Engquist}},
  \bibinfo{author}{\bibfnamefont{S.}~\bibnamefont{Osher}}, \bibnamefont{and}
  \bibinfo{author}{\bibfnamefont{S.~R.} \bibnamefont{Chakrabarty}},
  \bibinfo{journal}{J. Comput. Phys.} \textbf{\bibinfo{volume}{71}},
  \bibinfo{pages}{231} (\bibinfo{year}{1987}).

\bibitem[{\citenamefont{Bruegmann et~al.}(2008)\citenamefont{Bruegmann,
  Gonzalez, Hannam, Husa, and Sperhake}}]{Brugmann:2007zj}
\bibinfo{author}{\bibfnamefont{B.}~\bibnamefont{Bruegmann}},
  \bibinfo{author}{\bibfnamefont{J.~A.} \bibnamefont{Gonzalez}},
  \bibinfo{author}{\bibfnamefont{M.}~\bibnamefont{Hannam}},
  \bibinfo{author}{\bibfnamefont{S.}~\bibnamefont{Husa}}, \bibnamefont{and}
  \bibinfo{author}{\bibfnamefont{U.}~\bibnamefont{Sperhake}},
  \bibinfo{journal}{Phys. Rev. D} \textbf{\bibinfo{volume}{77}},
  \bibinfo{pages}{124047} (\bibinfo{year}{2008}).

\bibitem[{\citenamefont{Kreiss and Oliger}(1973)}]{Kreiss73}
\bibinfo{author}{\bibfnamefont{H.~O.} \bibnamefont{Kreiss}} \bibnamefont{and}
  \bibinfo{author}{\bibfnamefont{J.}~\bibnamefont{Oliger}},
  \emph{\bibinfo{title}{Methods for the approximate solution of time dependent
  problems}} (\bibinfo{publisher}{GARP publication series No. 10},
  \bibinfo{address}{Geneva}, \bibinfo{year}{1973}).

\bibitem[{\citenamefont{Gourgoulhon et~al.}(2001)\citenamefont{Gourgoulhon,
  Grandcl{\'e}ment, Taniguchi, Marck, and Bonazzola}}]{Gourgoulhon01}
\bibinfo{author}{\bibfnamefont{E.}~\bibnamefont{Gourgoulhon}},
  \bibinfo{author}{\bibfnamefont{P.}~\bibnamefont{Grandcl{\'e}ment}},
  \bibinfo{author}{\bibfnamefont{K.}~\bibnamefont{Taniguchi}},
  \bibinfo{author}{\bibfnamefont{J.~A.} \bibnamefont{Marck}}, \bibnamefont{and}
  \bibinfo{author}{\bibfnamefont{S.}~\bibnamefont{Bonazzola}},
  \bibinfo{journal}{Phys. Rev. D} \textbf{\bibinfo{volume}{63}},
  \bibinfo{pages}{064029} (\bibinfo{year}{2001}).

\bibitem[{\citenamefont{Taniguchi and Gourgoulhon}(2002)}]{Taniguchi02b}
\bibinfo{author}{\bibfnamefont{K.}~\bibnamefont{Taniguchi}} \bibnamefont{and}
  \bibinfo{author}{\bibfnamefont{E.}~\bibnamefont{Gourgoulhon}},
  \bibinfo{journal}{Phys. Rev. D} \textbf{\bibinfo{volume}{66}},
  \bibinfo{pages}{104019} (\bibinfo{year}{2002}).

\bibitem[{lor()}]{lorene}
\urlprefix\url{http://www.lorene.obspm.fr}.

\bibitem[{\citenamefont{Bildsten and Cutler}(1992)}]{Bildsten92}
\bibinfo{author}{\bibfnamefont{L.}~\bibnamefont{Bildsten}} \bibnamefont{and}
  \bibinfo{author}{\bibfnamefont{C.}~\bibnamefont{Cutler}},
  \bibinfo{journal}{Astrophys. J.} \textbf{\bibinfo{volume}{400}},
  \bibinfo{pages}{175} (\bibinfo{year}{1992}).

\bibitem[{\citenamefont{Baker et~al.}(2007{\natexlab{b}})\citenamefont{Baker,
  van Meter, McWilliams, Centrella, and Kelly}}]{Baker-etal:2007b}
\bibinfo{author}{\bibfnamefont{J.~G.} \bibnamefont{Baker}},
  \bibinfo{author}{\bibfnamefont{J.~R.} \bibnamefont{van Meter}},
  \bibinfo{author}{\bibfnamefont{S.~T.} \bibnamefont{McWilliams}},
  \bibinfo{author}{\bibfnamefont{J.}~\bibnamefont{Centrella}},
  \bibnamefont{and} \bibinfo{author}{\bibfnamefont{B.~J.} \bibnamefont{Kelly}},
  \bibinfo{journal}{Phys. Rev. Letters} \textbf{\bibinfo{volume}{99}},
  \bibinfo{pages}{181101} (\bibinfo{year}{2007}{\natexlab{b}}).

\bibitem[{\citenamefont{Boyle et~al.}(2007)\citenamefont{Boyle, Barrow, Kidder,
  {Mrou\'e}, Pfeiffer, Scheel, Cook, and Teukolsky}}]{Boyle:2007ft}
\bibinfo{author}{\bibfnamefont{M.}~\bibnamefont{Boyle}},
  \bibinfo{author}{\bibfnamefont{D.~A.} \bibnamefont{Barrow}},
  \bibinfo{author}{\bibfnamefont{L.~E.} \bibnamefont{Kidder}},
  \bibinfo{author}{\bibfnamefont{A.~H.} \bibnamefont{{Mrou\'e}}},
  \bibinfo{author}{\bibfnamefont{H.~P.} \bibnamefont{Pfeiffer}},
  \bibinfo{author}{\bibfnamefont{M.~A.} \bibnamefont{Scheel}},
  \bibinfo{author}{\bibfnamefont{G.~B.} \bibnamefont{Cook}}, \bibnamefont{and}
  \bibinfo{author}{\bibfnamefont{S.~A.} \bibnamefont{Teukolsky}},
  \bibinfo{journal}{Phys. Rev. D} \textbf{\bibinfo{volume}{76}},
  \bibinfo{pages}{124038} (\bibinfo{year}{2007}).

\bibitem[{\citenamefont{Shibata}(1999)}]{Shibata99c}
\bibinfo{author}{\bibfnamefont{M.}~\bibnamefont{Shibata}},
  \bibinfo{journal}{Phys. Rev. D} \textbf{\bibinfo{volume}{60}},
  \bibinfo{pages}{104052} (\bibinfo{year}{1999}).

\end{thebibliography}

\end{document}